\documentclass[useAMS,usegraphicx,usenatbib]{mn2e}

\usepackage{ulem}
\usepackage{color}
\usepackage{graphics,graphicx}
\usepackage{times} 
\usepackage{amssymb}
\usepackage{amsmath}
\usepackage{lscape}
\usepackage{url}
\usepackage{multirow}
\usepackage{ctable}
\usepackage{threeparttable}
\newif\ifAMStwofonts
\AMStwofontstrue
\def\xmm{{\it XMM-Newton}}

\def\suzaku{{\it Suzaku}}

\def\chandra{{\it Chandra}}

\def\bepposax{{\it BeppoSAX~\/}}

\def\epicpn{{\it EPIC}{\rm-pn}}
\def\epicmos1{{\it EPIC}{\rm-MOS1~\/}}
\def\epicmos2{{\it EPIC}{\rm-MOS2 ~\/}}
\def\epicmos{{\it EPIC}{\rm-MOS}}

\def\deg{$^{\circ}$}

\def\kmps{\hbox{$\rm\thinspace km~s^{-1}$}}

\def\H0{{\rm ~km~s^{-1}~Mpc^{-1}}}

\def\kev{\hbox{\rm keV}}

\def\ctps{\hbox{$\rm\thinspace ct~s^{-1}$}}

\def\atpcm{{\rm atom~cm$^{-2}$}}

\def\ergcmps{\hbox{\rm erg~cm~s$^{-1}$}}

\def\msun{\hbox{$\rm\thinspace M_{\odot}$}}

\def\chisq{{$\chi^{2}$}}
\def\rchi{{$\chi^{2}_{\nu}$~\/}}

\def\xspec{\hbox{\small XSPEC}}
\def\xspecv{\hbox{\small XSPEC}\, v12.6.0f}

\def\xmmselect{\hbox{\rm{\small XMMSELECT}}}
\def\ftool{\hbox{\rm{\small FTOOL}}}

\def\addspec{\hbox{\rm{\small ADDSPEC}}}

\def\grppha{\hbox{\rm{\small GRPPHA~\/}}}

\def\sas{\hbox{\rm{\small SAS~\/}}}
\def\epchain{\hbox{\rm{\small EPCHAIN}}}

\def\rmfgen{\hbox{\rm{\small RMFGEN}}}
\def\arfgen{\hbox{\rm{\small ARFGEN}}}

\def\epfast{\rm{\small EPFAST}}

\def\xstar{\hbox{\rm{\small XSTAR~\/}}}
\def\grid25{\hbox{\rm{\small GRID25}}}
\def\pl{\rm{\small POWERLAW}}

\def\laor{\rm{\small LAOR}}
\def\laortwo{\rm{\small LAOR2}}

\def\tbabs{\rm{\small TBABS}}
\def\diskbb{\rm{\small DISKBB}}

\def\refbhb{\rm{\small REFBHB}}

\def\reflionx{\rm{\small REFLIONX}}
\def\xillver{\rm{\small XILLVER}}
\def\cdid{\rm{\small CDID}}

\def\kdblur{\rm{\small KDBLUR}}

\def\kerrconv{\rm{\small KERRCONV}}

\def\comptt{\rm{\small COMPTT}}
\def\compps{\rm{\small COMPPS}}

\def\oviii{\hbox{\rm O\,{\small VIII}}}

\def\neii{\hbox{\rm Ne\,{\small II}}}
\def\neiii{\hbox{\rm Ne\,{\small III}}}
\def\ka{K\,$\alpha$}

\def\lya{Ly\,$\alpha$}

\def\eg{{\it e.g.~\/}}
\def\etc{{\it etc.}}
\def\ie{{\it i.e.~\/}}

\def\la{\mathrel{\hbox{\rlap{\hbox{\lower4pt\hbox{$\sim$}}}{\raise2pt\hbox{$<$}}}}}
\def\ga{\mathrel{\hbox{\rlap{\hbox{\lower4pt\hbox{$\sim$}}}{\raise2pt\hbox{$>$}}}}}

\def\d25{D$_{25}$}
\def\nh{{$N_{\rm H}$}}

\def\.25{0.25 keV\thinspace}

\def\mbh{\rm $M_{\rm BH}$}

\def\rg{$R_{\rm G}$}
\def\rin{$R_{\rm in}$}
\def\rout{$R_{\rm out}$}

\def\spin{\rm $cJ$/$GM^{2}$}

\def\xte{\rm XTE~J1650-500}
\def\mcg{\rm MCG--6-30-15}

\title[The Similarity of Broad Iron Lines in XRBs and AGNs]{The Similarity of Broad Iron Lines
in X-ray Binaries and Active Galactic Nuclei}

\author[D.\,J. Walton, R.\,C. Reis, E.\,M. Cackett, A.\,C. Fabian \& J.\,M. Miller]
{\parbox{7.in}{D.\,J. Walton$^{1}$ \thanks{E-mail: dwalton@ast.cam.ac.uk},
R.\,C. Reis$^{1,2}$,
E.\,M. Cackett$^{1}$,
A.\,C. Fabian$^{1}$ and
J.\,M. Miller$^{2}$ \\
\footnotesize
$^{1}$ \it{Institute of Astronomy, Cambridge University, Madingley Road, Cambridge, CB3 0HA} \\
$^{2}$ \it{Department of Astronomy, University of Michigan, 500 Church Street, Ann Arbor, MI 48109, USA}}}
\date{}

\begin{document}
\pagerange{\pageref{firstpage}--\pageref{lastpage}}
\pubyear{2010}
\maketitle
\label{firstpage}

\begin{abstract}We have compared the 2001 \xmm\ spectra of the stellar mass black hole binary \xte\
and the active galaxy \mcg, focusing on the broad, excess emission features at $\sim$4--7\,\kev\
displayed by both sources. Such features are frequently observed in both low mass X-ray binaries
and active galactic nuclei. For the former case it is generally accepted that the excess arises due
to iron emission, but there is some controversy over whether their width is partially enhanced by
instrumental processes, and hence also over the intrinsic broadening mechanism. Meanwhile, in the
latter case, the origin of this feature is still subject to debate; physically motivated reflection
and absorption interpretations are both able to reproduce the observed spectra. In this work we
make use of the contemporaneous \bepposax data to demonstrate that the breadth of the excess observed
in \xte\ is astrophysical rather than instrumental, and proceed to highlight the similarity of the
excesses present in this source and \mcg. Both optically thick accretion discs and optically thin coronae,
which in combination naturally give rise to relativistically-broadened iron lines when the disc
extends close to the black hole, are commonly observed in both class of object. The simplest solution
is that the broad emission features present arise from a common process, which we argue must be
reflection from the inner regions of an accretion disc around a rapidly rotating black hole; for
\xte\ we find spin constraints of $0.84 \leq a^{*} \leq 0.98$ at the 90 per cent confidence level. Other
interpretations proposed for AGN add potentially unnecessary complexities to the theoretical
framework of accretion in strong gravity.
\end{abstract}


\section{Introduction}

Simple models for the accretion of material onto a black hole (of any mass) usually
include two physically distinct components. The first is an optically thick, 
geometrically thin accretion disc (\citealt{Shakura73}) formed by material which
is gravitationally bound to the black hole, but has sufficient angular momentum to
avoid falling directly onto the black hole in the manner discussed by \cite{Bondi52}.
The second is a corona of hot electrons, analogous to that seen around the Sun
(although the exact geometry and energy distribution of these electrons is not yet
well determined, see \eg \citealt{eqpair}), populated by both thermal
and probably non-thermal electrons (\citealt{Malzac09}). The emission processes
relevant to each of these components are relatively well understood. Viscous
interactions within the disc heat the material and thermalise the emission, allowing
the disc to be treated as a series of black bodies of increasing temperature with
decreasing radius. Some fraction of the thermal emission from the disc is then
reprocessed by the coronal electrons, and Compton up-scattered into a powerlaw-like
emission component (\citealt{Haardt91}), with a high energy cut-off determined by
the temperature of the thermal electrons, and a low energy roll-over determined by the
temperature of the accretion disc.

The temperature at which the thermal emission from the disc is expected to be observed
depends on the mass of the black hole. For the black holes associated with X-ray
Binaries (XRBs; \mbh\ $\sim10$\,\msun) the discs are relatively hot and the thermal
emission should be observed in the X-ray band at $\sim$0.2--1.0\,\kev, while the discs
around the much larger black holes associated with active galactic nuclei (AGNs; \mbh\
$\gtrsim 10^{6}$ \msun) are much cooler and should be observed in the ultraviolet (UV).
However, in both cases the temperature of the coronal electrons is so high
($\sim$100\,\kev) that the Comptonised emission is always observed in X-rays.
Observations of both stellar and supermassive black holes in the UV and X-ray bands
have confirmed the presence of these emission components; for XRBs see recent reviews
by \cite{Remillard06rev} and \cite{Done07rev}, for AGNs see \cite{Elvis10sed} and
references therein.

With the onset of high-resolution CCD spectroscopy it has become clear that the broadband
X-ray spectra of AGN are in many cases far more complex than the simple powerlaw continuum
expected to be produced by the black hole corona. Additional features arise which can,
depending on the intrinsic continuum adopted, be interpreted as originating from either
emission or absorption processes, and viable, physically motivated models have been
proposed for each case. In emission interpretations the two most prominent features are a
broad and smooth excess below $\sim$1.0\,\kev, commonly referred to as
the `soft excess', and a second broad excess over $\sim$4.0--8.0\,\kev, while in
absorption models these are instead interpreted as a broad absorption trough spanning
$\sim$1.0--5.0\,\kev. The energy of these features, however interpreted, remains remarkably
constant, despite the observed AGN hosting black holes spanning many orders of magnitude in
mass (see \eg \citealt{Gierlin04}, \citealt{Crummy06}, \citealt{Nandra07},
\citealt{Miniutti09}, \citealt{Brenneman09}), which strongly suggests that if they are due to
additional emission they must also originate through atomic processes (absorption being an
atomic process by nature).

One of the successful emission models is the disc reflection/light bending interpretation
which assumes that the corona is compact (or centrally concentrated) and emits some fraction
of its flux towards the accretion disc. This is reprocessed into an additional `reflected'
emission component via backscattering, fluorescence, and Compton scattering interactions
within the surface of the disc (\citealt{George91}). This reflection component includes
discrete atomic features imprinted by the disc, the most prominent of which is usually the
neutral iron \ka\ emission line at 6.4\,\kev. However, the strong gravity close to a black
hole broadens and skews these features via Doppler shifts, beaming and gravitational redshifts
(\citealt{Fabian89, kdblur}). The broad excess over $\sim$4.0--8.0\,\kev\ is thus attributed
to the iron \ka\ emission line broadened by such relativistic effects, and the large number
of emission lines from iron and lower-mass elements in the $\sim$0.5--1.0\,\kev\ range are
blended together into a broad and smooth emission component, similar to that observed. The
strong gravity also focuses some additional amount of the hard emission down onto the disc
through gravitational light bending. This phenomenon causes the spectral variability often
observed; if the region in which the Comptonised continuum is produced varies in distance
from the central black hole, the fraction of the coronal flux focussed onto the disc, and
hence that observed, also varies (see \eg \citealt{lightbending}). For a recent review on
X-ray reflection, see \cite{Fabian10rev}.

An alternative interpretation is that the observed deviations from the powerlaw continuum
arise as a result of complex absorption due to structures of intervening clouds and
infalling/outflowing material. Combinations of these structures can lead to complex
combinations of neutral, partially and fully ionised material. Indeed, in a number of
AGN there are clear, undeniable signatures of absorption due to partially ionised
material (see \eg \citealt{Blustin05}); these `warm' absorbers are usually associated
with mildly outflowing material from the dusty torus or clouds in the broad line region.
In addition, there are also detections of absorption by highly ionised material with
mildly relativistic blueshifts claimed, which are interpreted as evidence for outflowing
disc winds (see \citealt{Tombesi10b}, \citealt{Reeves09}, \citealt{Pounds09}, and references
therein), although it may be necessary to treat a number of these claims with caution (see
\citealt{Vaughan08}). However, in order to reconcile the large extent of the apparent absorption
trough in energy (up to $\sim$5\,\kev) with its relatively shallow appearance, it is usually
necessary to invoke absorption by partially ionised material that only covers some fraction
of the X-ray source. With appropriate combinations of these various types of absorption
the observed spectra can be reproduced, and spectral variability may be explained via
changes in the covering fraction, and/or column density and ionisation of the absorbing
material. For a review of the absorption processes relevant to AGN see \cite{Turner09rev}.

Even with the quality and quantity of data provided by the latest generation of X-ray
observatory, such as \xmm\ and \chandra, it has proven very difficult to statistically
distinguish between these interpretations for AGN, a case in point being the highly
variable AGN \mcg\ (see \citealt{Fabian02MCG} and \citealt{Miller08}, who construct
statistically acceptable reflection and absorption dominated models respectively). In
principle, the simultaneous coverage of the $\sim$0.5--50.0\,\kev\ energy range possible
with the \suzaku\ satellite may be able to distinguish between disc reflection and pure
absorption interpretations in some cases. Reflection models predict the presence of an
additional, broad emission feature at $\sim$20--30\,\kev, referred to as the Compton Hump
as it arises due to the interplay of Compton up-scattering of low energy photons and
photoelectric absorption of high energy photons within the reflecting medium. Features
consistent with the Compton hump are fairly commonly seen in AGN (see \eg
\citealt{Walton10Hex}). However, owing to the relatively poor quality data currently
available above 10\,\kev\ in many cases, these high energy features can also be
consistent with the presence of a partially covering, Compton thick absorber (see \eg
\citealt{Turner09, Reeves09}).

One of the strongest individual cases in favour of the reflection interpretation is the
narrow-line Seyfert 1 galaxy 1H\,0707-495, in which reverberation between the powerlaw
and reflection components has been detected (see \citealt{FabZog09}, \citealt{Zoghbi10}).
\cite{LMiller10} argue such apparent reverberation can still be reproduced by distant
reflection from the winds required by an absorption dominated interpretation. However, no
absorption model that simultaneously reproduces the reverberation \textit{and} the
observed energy spectrum is offered. In addition, it is worth noting that soft excesses
are observed in AGN which display no obvious indication of any absorption over that
expected due to the Galactic interstellar medium (ISM), \eg Ark\,120 (\citealt{Vaughan04,
Nardini11}) and Ton\,S180 (\citealt{Vaughan02}).

Interestingly, high quality X-ray spectroscopy of XRBs has also revealed the presence of
broad excesses over the Comptonised continuum in the $\sim$4.0--8.0\,\kev\ energy range,
which are again frequently interpreted as relativistically broadened iron emission lines,
\eg \xte\ (\citealt{Miller02b, Miniutti04}), GX~339-4 (\citealt{Reis08gx}), GRS~1915+105
(\citealt{blum09}), Swift\,J1753.5-0127 (\citealt{Reis09spin}), XTE\,J1752-223
(\citealt{Reis1752}). In addition, high energy features consistent with the Compton Hump
are also frequently observed in BHBs (see \eg \citealt{Reis10lhs}), supporting the theory
that X-ray reflection is present in these sources. Studying XRBs has the advantage that
they are often brighter in flux than their AGN counterparts, so the presence of absorption
due to outflowing winds, \etc\ can be tested in greater detail, and their contribution,
if any, to the observed spectrum determined with much greater confidence.

However, the presence of relativistically broadened iron lines in XRBs is not without its own
controversy. Although it is generally accepted that these features do arise due to iron \ka\
emission, and that this emission has some finite width, there is some debate over how broad
the lines actually are, in particular whether instrumental effects act to artificially enhance
the apparent width and skew of the line profile, and therefore over the mechanism that
broadens them. This is entwined with the debate over whether the accretion disc extends in
to the inner-most stable circular orbit (ISCO). While it is generally accepted that this is
the case at high accretion rates, and that at some low accretion rate the optically thick disc
most likely begins to truncate at a larger radius within which a hot, advection dominated
accretion flow arises, the accretion rate at which truncation occurs is still under dispute.
Some authors, \eg \cite{Done07rev}, argue that the disc begins to truncate during the
transition between the high/soft and low/hard accretion states, while others, \eg
\cite{Reis10lhs}, argue that the disc remains at the ISCO until much lower accretion rates
(for a review of the `standard' accretion states displayed by BHBs see \citealt{Remillard06rev}).

If, as is likely, the line is produced in the disc, then as the disc recedes the line profile
should evolve and become narrower. However, BHBs in the low/hard state still appear to display
similarly broad excesses to other accretion states (see \eg \citealt{Reis1752}). \cite{Liu11}
proposed a modification to the truncated disc model, in which the outer disc truncates, but
some inner portion remains, resulting in a disc with a hole in it. Such a geometry would allow
the truncated disc model and the broad iron line observations to be reconciled without invoking
a separate medium that produces the line. However, through analysis of the variability displayed
by both the thermal and Comptonised components of GX\,339-4 in the low/hard state, \cite{Wilkinson09}
show that in order for the variability to be driven by fluctuations in the mass accretion rate,
as is generally thought to be the case, the disc must extend uninterrupted to at least 20\,\rg,
placing strict constraints on the location and size of any hole which render this scenario
unlikely. The truncated disc interpretation therefore requires that the dominant line broadening
mechanism is some process other than Doppler/gravitational effects, as this interpretation
frequently requires that the disc extends close to the black hole.

An alternative mechanism that has been invoked to explain broad line profiles is Compton
broadening, in which Compton scattering of line photons modifies their energy (see \eg
\citealt{Czerny91}, \citealt{Misra99}, \citealt{diSalvo05}, \citealt{Titarchuk09}). This
process broadens and, depending on the conditions of the Comptonising region, can even skew
the line profile. Some contribution from Compton broadening is unavoidable for any realistic
method of line production, as the line emission will necessarily be Compton scattered as it
escapes from the medium in which it is produced. If the line is associated with the disc,
Compton broadening is expected to be more significant process for BHBs than for their AGN
counterparts owing to their relative disc temperatures (\citealt{refbhb}).
The challenge then becomes determining the relative contribution to the line profile of this
process and the relativistic effects discussed above. However, it has also been proposed that
the line does not originate in the disc, but is instead produced via recombination in the
corona or in a disc wind. These scenarios require Compton broadening to play a prominent
role in determining the line profile. We note that, given the changes in the corona and the
changes in mass inflow and outflow rate typically observed during the evolution of BHB
outbursts, evolution in the line profile would also be expected for these latter scenarios.

Here, we present an analysis of the high quality X-ray spectra of the XRB \xte, which
displays a broad line profile. Both relativistic and Compton broadening are examined,
and we demonstrate that the former scenario is strongly preferred. We then proceed to
compare the high energy spectra of \xte\ and the AGN \mcg, and demonstrate the
similarity of the features identified as broad iron lines in each case. This work is
structured as follows: section \ref{sec_red} gives details on the data reduction,
section \ref{sec_spec} presents our spectral analysis, highlighting the similarity of
the two line profiles, and finally section \ref{sec_disc} presents our discussion, in
which we use this comparison as the basis for a simple, logical argument in favour of
the reflection interpretation for AGN.

\section{Observations and Data Reduction}
\label{sec_red}

\xte\ and \mcg\ were both observed with \xmm\ (\citealt{XMM}) in 2001 September and
August respectively. In addition, \xte\ was observed three times by \bepposax
(\citealt{SAX}) during the same outburst. The first \bepposax and the \xmm\
observations of \xte, separated by about a day, caught the source in a rising
intermediate state, \ie during the transition between the low/hard and high/soft states
in which the thermal and Comptonised emission components are both strong. The latter two
\bepposax observations, taken further along the evolution of the outburst, found the
source in a more traditional high/soft state. These observations were selected because
they represent some of the best quality X-ray data available for XRBs and AGNs, as the
data obtained have good photon statistics, the \xmm\ observation provides low energy
coverage and the \bepposax data provides a view of the emission from \xte\ at
$\sim$6\,\kev\ with an independent detector. Here we detail the data reduction procedure
adopted for each source. In both cases the \xmm\ reduction was carried out with the \xmm\
Science Analysis System (\sas v10.0.0).

\subsection{XTE~J1650-500}
\label{sec_xte_red}

The 2001 observation of \xte\ was taken on September 13th with the \epicpn\
(\citealt{XMM_PN}) CCD in burst mode, so despite its extremely high flux, the spectrum
obtained with this detector is not expected to suffer from pile-up\footnote{Pile-up
refers to the situation in which more than one photon is incident on a CCD pixel within
a single readout time. These are incorrectly registered as a single photon at an
artificially high energy. For a detailed description of this effect see
\cite{Miller_pileup} and references therein.}. The data reduction procedure adopted for
this source follows closely that outlined in \cite{epicpn_burst}. The \epicpn\
observation data files were processed using \epchain\ to produce calibrated event lists
from which spectral and timing products may be extracted. Burst mode is similar to
Timing mode and operates with extremely fast readout times; 200 lines of CCD pixels are
fast-shifted in 14.4 $\mu$s as the source data is being recorded. This readout process
leads to loss of spatial information in the shift (RAWY/CCD line number) direction,
such information is only available in the direction on the CCD perpendicular to the
readout (RAWX/CCD column number). Despite the extremely fast readout employed in
burst mode, \cite{epicpn_burst} found that events with RAWY $>$ 140 contain piled-up
data, so we extracted the source spectrum from the region bounded by $0 \leq$ RAWY
$\leq 140$ and $27 \leq$ RAWX $\leq 46$. In addition, we also only included single
and double events, and excluded border pixel events. As \xte\ was so
bright during this observation, there was no region on the operational CCD free of
source counts, so we did not extract a background spectrum following the recommendation
of the \xmm\ burst mode reduction guide\footnote{http://xmm.esac.esa.int/sas/current/documentation/threads/}.
Given the high source countrate (see below), the contribution of background counts
should be negligible. The redistribution matrices and auxiliary response files were
generated with \rmfgen\ and \arfgen\ respectively. The spectrum obtained had a good
exposure time of 685\,s, and an average total countrate of 2504\,$\pm$\,2\,\ctps.
Finally, the spectrum was grouped using \grppha\ such that each energy bin contained
a minimum of 25 counts, so that the probability distribution of counts within each bin
can be considered Gaussian, and hence the use of the $\chi^2$ statistic is appropriate
when performing spectral fits. The \epicmos\ detectors were not operated for this
observation, and we do not consider the spectrum obtained with the RGS instrument in
this work, as it suffers quite severely from pile-up effects.

Throughout this work, we also make use of the three \bepposax observations taken
during the course of the same outburst. These were obtained on 2001 September
11th, 21st and October 3rd respectively. We reduced the MECS (\citealt{SAX_MECS})
data for each observation following the prescription outlined in the \bepposax
cookbook\footnote{http://heasarc.gsfc.nasa.gov/docs/sax/abc/saxabc/saxabc.html}.
In each case we defined a circular source region of radius 8', and a circular
background region free of other sources of the same size, and use the appropriate
September 1997 auxiliary response and redistribution files. The good exposure
times and average countrates obtained for each observation are 47, 68 and 24\,ks
and 34.18\,$\pm$\,0.03, 42.18\,$\pm$\,0.03 and 32.57\,$\pm$\,0.03\,\ctps\
respectively. Note that the MECS detector has a lower energy resolution than
\epicpn\ ($\sim$750\,eV in comparison to $\sim$150\,eV\ at 6\,\kev).

\subsection{MCG--6-30-15}

The long 2001 observation of \mcg\ was taken in small window mode, and spanned three
orbits. Data reduction was carried out separately for each orbit, largely according
to the standard prescription provided in the online guide\footnote{http://xmm.esac.esa.int/}.
In this work we focus on the \epicpn\ data for direct comparison with \xte. The
observation data files were processed using \epchain\ to produce calibrated event lists.
Spectra were produced for the 0.5--10.0\,\kev\ energy range selecting only single and
double events using \xmmselect, and periods of high background were treated according
to the method outlined by \cite{Picon04}, with the signal-to-noise ratio maximised for
the 6.0--10.0 \kev\ energy band, where the neutral and ionised iron transitions are
expected to be observed. A circular source region of radius 37'' was defined in order
to include as many source counts as possible without including CCD chip edges, and a
larger circular background region of radius 54'' was chosen in an area of the same CCD
free of other sources. The redistribution matrices and auxiliary response files were
generated with \rmfgen\ and \arfgen. After performing the data reduction separately for
each of the orbits, the spectra were combined using the
\ftool\footnote{http://heasarc.nasa.gov/ftools/ftools\_menu.html} \addspec, which
combines spectra in a response weighted manner to account for any differences there may
be in \eg the exposure times, \etc\ \addspec\ also automatically combines the
instrumental responses and background spectra associated with the source spectra. The
resulting spectrum had a good exposure time of 221\,ks, and an average total count
rate of 28.15\,$\pm$\,0.01\,\ctps. Finally, for the same reason as for \xte, the spectrum
was grouped using \grppha to have a minimum of 25 counts in each energy bin.

\section{Spectral Analysis}
\label{sec_spec}

Here we detail our spectral analysis of \xte\ and \mcg. The approach for \xte\ is
simple: we begin by modelling the continuum to highlight the broad line profile,
before modelling the line in a phenomenological fashion and finally moving on to
a more physical treatment of the disc reflection. In addition, we also consider
Compton broadening in some detail and find that strong relativistic effects are
still required to explain the breadth of the line. We then construct a simple
phenomenological model for the spectrum of \mcg\ in order to again highlight the
line profile in this source, and its similarity to that of \xte\ (a full physical
interpretation of \mcg\ is presented in \citealt{Chiang11}, utilising all currently
available data, and so will not be undertaken here). Spectral analysis is performed
with \xspecv\ (\citealt{XSPEC}), and quoted uncertainties on model parameters are
the 90 per cent confidence limits for one parameter of interest determined by
\chisq\ variation, unless stated otherwise. In both cases, due to calibration
uncertainties possibly related to a deficient charge transfer inefficiency (CTI)
correction at high count rates (see section \ref{sec_inst} for a description of the
effects of CTI), significant negative residuals were continually observed associated
with the instrumental silicon K and gold M edges at $\sim$1.8 and 2.2\,\kev\
respectively, hence the energy range 1.7--2.5\,\kev\ has been excluded from our
analysis throughout this work. We adopt the solar abundances given in \cite{tbabs}.

\subsection{XTE\,J1650-500}
\label{sec_xte}

\subsubsection{Continuum Modelling}
\label{sec_xte_cont}

\begin{table*}
  \caption{Key continuum and iron line parameters obtained by phenomenologically
modelling the broad line profiles of \xte\ and \mcg\ with \laor\ line models.
Parameters marked with * have not been allowed to vary, and those with `p' quoted
as one of the limits have reached the relevant constraint imposed upon it. The rest
frame energies of the iron emission lines were required to be in the range
6.40--6.97\,\kev. The temperatures quoted are of the \diskbb\ component for \xte,
and of the black body included to account for the soft excess in \mcg.}
\begin{center}
\begin{tabular}{c c c c c c c c c c c c}
\hline
\hline
\\[-0.25cm]
& $\Gamma$ & $T$ & $E$ & $R_{\rm in}$ & $R_{\rm br}$ & $R_{\rm out}$ & $q_{\rm inner}$ & $q_{\rm outer}$ & $i$ & $EW$ \\
\\[-0.3cm]
& & (\kev) & (\kev) & (\rg) & (\rg) & (\rg) & & & (deg) & (eV) \\
\\[-0.25cm]
\hline
\\[-0.25cm]
\xte & $2.14^{+0.02}_{-0.01}$ & $0.295 \pm 0.007$ & $6.97^{p}_{-0.30}$ & $1.47^{+0.08}_{-0.03}$ & 6* & 400* & $>8.2$ & 3* & $69^{+3}_{-8}$ & $370^{+100}_{-70}$ \\
\\[-0.2cm]
\mcg & $1.961 \pm 0.004$ & $0.123 \pm 0.002$ & $6.40^{+0.04}_{p}$ & $1.9^{+0.2}_{-0.1}$ & $8^{+28}_{-3}$ & 400* & $4.8^{+1.6}_{-0.9}$ & $2.2 \pm 0.3$ & $47^{+4}_{-2}$ & $430^{+120}_{-70}$ \\
\\[-0.25cm]
\hline
\hline
\end{tabular}
\label{tab_laor}
\end{center}
\end{table*}

We model the continuum of \xte\ with the standard black hole binary (BHB) components,
namely a multi-colour black body accretion disc (modelled with \diskbb; \citealt{diskbb})
which accounts for the thermal emission from the disc and dominates at soft energies,
and a powerlaw, associated with Comptonisation of the thermal emission by electrons
in a corona, which dominates at hard energies. Both of these are modified by Galactic
absorption, accounted for with the \tbabs\ absorption code (\citealt{tbabs}). The
column density of the ISM, as well as its Oxygen, Neon and iron abundances, were allowed
to vary; all other elements were assumed to have solar abundances. The 4.0--8.0\,\kev\
data were not considered at this stage of the analysis, in order to minimize any
contribution from the iron emission region to our determination of the continuum. This
model was fit to the remaining data, and although statistically it is not formally
acceptable, with \rchi = 1184/865, it provides a good representation of the general
shape of the spectrum. The relatively poor fit statistic is mostly due to residual
features at $\sim$0.6--0.8\,\kev\ (see Fig. \ref{fig_xte_line}), which will be discussed
in section \ref{sec_soft}; the data closer to the 4.0--8.0~\kev\ range, where the iron
line is expected to be observed, is well modelled.

\subsubsection{Line Profile}
\label{sec_xte_line}

\begin{figure}
\begin{center}
\rotatebox{0}{
{\includegraphics[width=230pt]{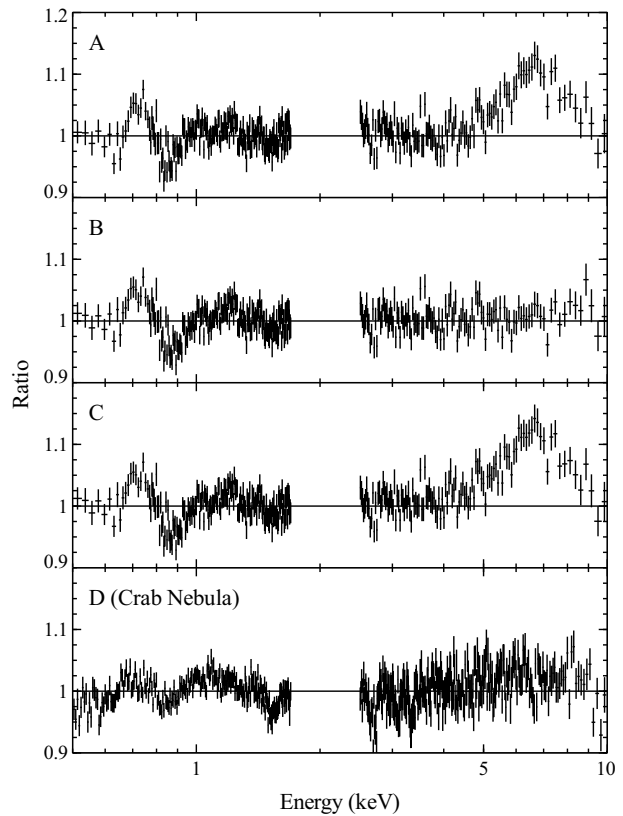}}
}
\end{center}
\caption{Data/model ratio plots for \xte. \textit{Panel A}: ratio plot to the continuum
model fit to the 0.5--4.0 and 8.0--10.0\,\kev\ data (see section \ref{sec_xte_cont}); the
broad iron line profile is clearly shown. \textit{Panel B}: the ratio plot when a
\laortwo\ line is included in the model; the iron line is clearly well modelled.
\textit{Panel C}: ratio plot to the model including the \laortwo\ line with the line
normalisation set to zero, to demonstrate that this component accounts for the hard
residuals without drastically altering the continuum. Soft residual features over the
$\sim$0.6--0.8\,\kev\ energy range are seen throughout, these are discussed in section
\ref{sec_refl} and section \ref{sec_abs}. \textit{Panel D}: ratio plot of the 2002 burst mode
spectrum of the Crab Nebula modelled with an absorbed powerlaw (see section \ref{sec_refl}).
Similar soft residuals are seen, implying these features possibly arise due to systematic
uncertainties. The spectra have been rebinned for clarity.}
\label{fig_xte_line}
\end{figure}

We now include the 4.0--8.0\,\kev\ data in our analysis. Fig. \ref{fig_xte_line} (panel
A) shows the data/model ratio of the full 0.5--10.0\,\kev\ spectrum for \xte\ to the
previously determined continuum model. It is clear that there is a large, broad excess
in the ratio spectrum in this energy range, which we interpret as arising due to iron
emission. Initially, we modelled the line profile with a Gaussian emission line
component, with the rest-frame energy of the line constrained to be between the neutral
iron \ka\ transition at 6.4\,\kev\ and the iron \lya\ transition at 6.97\,\kev, as the
discs in XRBs are likely to be highly ionised (\citealt{refbhb}). With the inclusion
of the Gaussian line the model provides a fairly good fit to the data, with \rchi =
1936/1663. The soft residuals highlighted previously are still present, but the line
profile is well modelled. We find that the line energy is constrained to the lower end
of the allowed energy range in contrast with the expectation for an ionised disc, with
$E_{\rm Gauss} <$ 6.5\,\kev, and the profile is very broad, with $\sigma =
1.17^{+0.24}_{-0.21}$\,\kev. The equivalent width obtained is $EW = 350^{+130}_{-110}$
eV.

If the iron line is primarily broadened by gravitational and Doppler effects, then
its width implies the disc must extend to the regions of strong gravity close to the
black hole. We replaced the Gaussian emission line with a \laortwo\ line
(\citealt{kdblur}), which applies the relevant relativistic effects to a Gaussian
emission line profile. The parameters of this component are the energy of the emission
line, the inner and outer radii of the accretion disc in gravitational radii
(\rg $=GM/c^2$), the inclination of the disc, $i$, the indices of the radial emissivity
profile, assumed to be a broken powerlaw of form:

\vspace{-0.35cm}
\begin{equation}
\epsilon(r) \propto \left\{\begin{matrix}
r^{-q_{\rm inner}} & r < R_{\rm br} \\ 
\\[-0.3cm]
r^{-q_{\rm outer}} & r > R_{\rm br}
\end{matrix} \right.
\label{eqn_bpl}
\end{equation}

\noindent{and the break radius of the emissivity profile, also in gravitational
radii. Recent simulations have suggested the radial emissivity profile may indeed
be more complex than a single powerlaw, showing a break at $\sim$6\,\rg\
(\citealt{Wilkins11, lightbending}). The outer radius of the disc was taken as
the maximum allowed by the model, 400\,\rg\ (this parameter is rarely well
constrained as even in the case of a Schwarzschild black hole the majority of the
emission arises from the innermost regions of the disc). The rest of the line
parameters were initially free to vary, although in the course of the modelling we
found that $R_{\rm br}$ and $q_{\rm outer}$ could not be constrained. These were
respectively set to $R_{\rm br}=6$\,\rg, after which the metric tends asymptotically
towards the Schwarzschild case even for a maximally rotating Kerr black hole, and
$q_{\rm outer}=3$, as expected for a lamp-post illumination model. With the
inclusion of the \laortwo\ line, the model again provides a fairly good fit to the
data, with \rchi = 1936/1661. The parameters and equivalent width ($EW$) obtained
for the \laortwo\ profile are given in Table \ref{tab_laor} and the data/model ratio
plot is shown in Fig. \ref{fig_xte_line} (panel B). The energy of 6.97\,\kev\
obtained for the iron line with this profile is more consistent with the expectation
of a highly ionised disc. In Fig. \ref{fig_xte_line} (panel C) we also show the
ratio plot for the model including the \laortwo\ line, but with the line
normalisation set to zero, to demonstrate that this component models the hard
residuals well without significantly altering the continuum model determined
previously.}

\subsubsection{High Energy Comptonisation}
\label{sec_compt}

\begin{figure}
\begin{center}
\rotatebox{0}{
{\includegraphics[width=230pt]{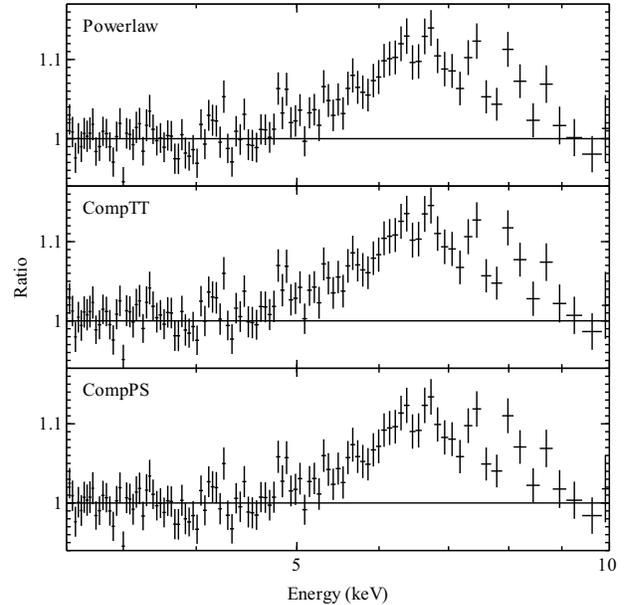}}
}
\end{center}
\caption{Data/model ratio plots for the \xmm\ observation of \xte, focussing on
the iron emission, after making use of various models for the high energy
Comptonised continuum. The models investigated are a simple powerlaw (top panel),
and the \comptt\ and \compps\ Comptonisation codes (middle and bottom panels
respectively). The choice of model does not effect the line profile obtained.}
\label{fig_cont_comp}
\end{figure}

Although the high energy Comptonised continuum is often powerlaw-like in the \xmm\
bandpass, we also investigated whether this approximation has any effect on the
line profile. To test this, we also modelled the high energy continuum with the
more physical Comptonisation codes \comptt\ (\citealt{comptt}) and \compps\
(\citealt{compps}). In both cases the temperature of the seed photon spectrum
was set to that of the \diskbb\ component, and we assumed a standard slab geometry.
These models were applied in the same manner detailed previously (\ie initially
excluding the 4.0--8.0\,\kev\ data), and the line profiles obtained examined. As
shown in Fig. \ref{fig_cont_comp}, use of these more sophisticated Comptonisation
models does not lead to any modification of the line profile. With the addition of
a \laortwo\ line we obtain fits of equivalent quality to the case using the simpler
powerlaw continuum in both cases (\rchi = 1933/1660 and 1941/1660 respectively).
The parameters obtained for the line profile with each of the models are given in
Table \ref{tab_compt}, and show excellent agreement within their statistical
uncertainties. Therefore, we conclude that a simple powerlaw component provides an
adequate representation of the Comptonised continuum in the bandpass considered.
We will proceed with the use of this approximation throughout the rest of this work.
In addition, it is clear that any systematic uncertainties introduced through our
choice of continuum model are negligible in comparison to the statistical parameter
uncertainties obtained.

\begin{table}
  \caption{Key iron line parameters obtained when modelling Comptonised continuum
with the \comptt\ and \compps\ codes. The line profiles have again been modelled with
a \laortwo\ component, in the same manner as with the simple powerlaw continuum. The
values obtained show excellent agreement with those presented in Table \ref{tab_laor}).}
\begin{center}
\begin{tabular}{c c c c c c}
\hline
\hline
\\[-0.25cm]
Model & & $E$ & \rin & $q_{\rm inner}$ & $i$ \\
\\[-0.3cm]
& & (\kev) & (\rg) & & (deg) \\
\\[-0.25cm]
\hline
\\[-0.25cm]
\comptt & & $6.97^{p}_{-0.32}$ & $1.5 \pm 0.1$ & $>6.6$ & $69^{+2}_{-4}$ \\
\\[-0.2cm]
\compps & & $6.97^{p}_{-0.44}$ & $1.5 \pm 0.1$ & $>8.7$ & $69^{+3}_{-7}$ \\
\\[-0.25cm]
\hline
\hline
\end{tabular}
\label{tab_compt}
\end{center}
\end{table}

\subsubsection{Disc Reflection}
\label{sec_refl}

\begin{figure}
\begin{center}
\rotatebox{270}{
{\includegraphics[width=160pt]{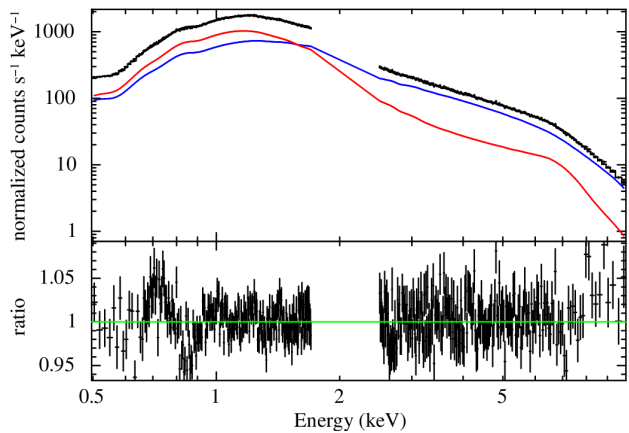}}
}
\end{center}
\caption{The basic disc reflection model (Model 1) for \xte\ presented in \S
\ref{sec_refl}, and the relative contributions of the components included. The
solid black line shows the total model, while the blue shows the powerlaw
continuum and the red shows the blurred reflection component (which includes
the thermal contribution from the disc). The spectra have been rebinned for
clarity.}
\label{fig_1650}
\end{figure}

We now embark upon a more physical consideration of the relativistic disc
reflection scenario for \xte, and replace the \diskbb\ and \laor\ components
with the \refbhb\ model (\citealt{refbhb}) convolved with the \kdblur\
kernel, which applies the same relativistic effects included in the \laor\
line model to any general input spectrum, but assumes a simpler single
powerlaw emissivity profile of form $\epsilon(r) \propto r^{-q}$. \refbhb\
is a self consistent reflection model which intrinsically calculates the iron
K-shell absorption and emission in the reflecting medium, correctly including
effects such as Compton broadening, which, as emphasised previously, is an
important process in BHBs due to the high temperature of the disc
(\citealt{refbhb}). It is calculated specifically for use with spectra from
BHBs as it includes the thermal emission from the accretion disc. Its key
parameters are the hydrogen number density and temperature of the accretion
disc ($n_{\rm H}$ and $kT$), the photon index of the illuminating powerlaw
continuum, the relative strength of the illumination with respect to the
thermal emission at the surface of the disc ($F_{\rm il}/F_{\rm th}$), and
the redshift of the reflecting medium. We assume the redshift is zero, while
the photon index of the illuminating emission was required to be the same as
that of the powerlaw continuum component. This model provides a fairly good
representation of the data, with \rchi = 1832/1663, and the iron line profile
is well modelled. The key parameters obtained are given in Table \ref{tab_refl}
as Model 1, and the relative contributions of the components are shown in Fig.
\ref{fig_1650}. As suggested by the results obtained with the \laortwo\ line
profile in section \ref{sec_xte_line}, we again find that the inner radius of
the disc is small, with \rin\ $\sim 2$\,\rg, although the radius obtained here
is slightly larger than with the simple \laortwo\ profile, owing to the self
consistent inclusion of Compton broadening in \refbhb. In addition, the value
obtained for $F_{\rm il}$/$F_{\rm th}$ suggests that the Comptonised emission
contributes significantly more to the total flux than the thermal emission
from the disc, which is not uncommon for a source during the transition from
the low/hard to the high/soft state (\citealt{Remillard06rev}).

In order to assess whether our choice of reflection model has any major
influence over the key parameter values obtained, we replaced \refbhb\ with
\reflionx\ (\citealt{reflion}), another self consistent reflection code. 
In particular, we are interested in the inner radius of the disc, which in turn
provides information on the spin of the black hole and is key in determining
the width of the line. The key parameters of \reflionx\ are the iron abundance
and ionisation parameter (defined as $\xi = L/nR^2$, where $L$ is the ionising
luminosity of the source, $n$ is the density of the absorber and $R$ is its
distance from the source) of the reflecting medium, and the photon index of the
ionising continuum. We again required the photon index parameters of both the
reflection model and the powerlaw continuum to be the same, and the iron
abundance was set to the solar value. As expected, the reflecting medium is
found to be highly ionised, with $\log \xi = 2.8$. However, although \reflionx\
includes treatment of all the same processes as \refbhb, it is calculated for
use with X-ray spectra from active galaxies, and does not include the direct
thermal emission from the disc in the correct fashion for BHBs. In order to
account for this emission we again included a \diskbb\ component. This
alternative reflection model is also able to reproduce the data fairly well,
with \rchi = 1848/1662. The inner radius obtained for the disc here is \rin\ =
$1.31^{+0.13}_{-0.07}$, even smaller than in the \refbhb\ case. This decrease
can be understood in terms of the differing situations for which the two
reflection models are calculated. We stress again that \reflionx\ is calculated
for use with active galaxies, and therefore assumes a much cooler temperature
and lower density for the accretion disc. This in turn reduces the contribution
from Compton broadening, hence additional gravitational broadening is required
to reproduce the line profile, and a smaller inner radius is obtained. For this
reason, the inner radius obtained with \refbhb\ is more appropriate as it
includes the correct amount of Compton broadening for XRBs.

It is worth briefly noting that the neutral column density for the ISM obtained
is consistently larger than that presented in some of the previous work on this source
(\eg \citealt{Miller04}, \citealt{Miller09}), possibly owing to the inclusion of
data in the range 0.5--0.7\,\kev\ in our analysis. If we restrict ourselves to a
consideration of the same 0.7--10.0\,\kev\ energy range analysed in
\cite{Miller09}, we obtain equivalently acceptable fits with a column density
consistent with theirs. However, in doing so the abundances of Oxygen, Neon and
iron obtained are all $\sim$2 times solar. The difference in column obtained
is therefore most likely to be due to the updated absorption models, solar
abundances and cross-sections used in this work. In addition, none of the models
presented required the energy of the neutral Oxygen recombination edge to be
shifted away from the expected value, as originally found in \cite{Miller02b}.

\subsubsection{Soft Residuals}
\label{sec_soft}

\begin{table*}
  \caption{Key Parameters obtained for the various models constructed using a reflection
interpretation for the \xmm\ spectrum of \xte\ (see text); parameters marked with * have
not been allowed to vary. Note that when included, the \neii\ and \neiii\ lines have been
required to contribute equal amounts to the negative residuals at $\sim$0.85\,\kev. Column
densities are given in $10^{21}~$\atpcm, abundances are quoted relative to the solar
abundances, equivalent widths are given in eV, radii in gravitational radii, inclinations
in degrees, temperatures in \kev\ and hydrogen number densities are given in $10^{20}$
atom cm$^{-3}$. For the model including the \xstar\ grid, the ionisation parameter is given
in \ergcmps, and the outflow velocity in \kmps. Note that these results are obtained with
unmodified \xmm\ data; for results obtained after the application of the \epfast\ tool see
section \ref{sec_inst}.}
\begin{center}
\begin{tabular}{c p{0.2cm} c p{0.2cm} c p{0.2cm} c p{0.2cm} c p{0.2cm} c p{0.2cm} c}
\hline
\hline
\\[-0.3cm]
Component & & Parameter & & \multicolumn{9}{c}{Value} \\
\\[-0.2cm]
& & & & Model 1 & & Model 2 & & Model 3 & & Model 4 & & Model 5 \\
\\[-0.3cm]
\hline
\hline
\\[-0.25cm]
\tbabs & & \nh & & $7.17^{+0.08}_{-0.09}$ & & $7.2^{+0.2}_{-0.1}$ & & $7.4^{+0.1}_{-0.2}$ & & $7.2^{+0.2}_{-0.1}$ & & $7.2^{+0.2}_{-0.1}$ \\
\\[-0.2cm]
& & $A_{\rm O}$ & &  & & $1.06^{+0.05}_{-0.02}$ & & $1.26^{+0.06}_{-0.04}$ & & $1.16 \pm 0.07$ & & $1.07^{+0.04}_{-0.06}$\\
\\[-0.2cm]
& & $A_{\rm Ne}$ & & $1.36^{+0.06}_{-0.03}$ & & $1.5 \pm 0.1$ & & $0.9^{+0.02}_{-0.3}$ & & $1.4 \pm 0.2$ & & $1.3 \pm 0.2$\\
\\[-0.2cm]
& & $A_{\rm Fe}$ & & $0.95^{+0.04}_{-0.07}$ & & $0.8^{+0.2}_{-0.1}$ & & $0.9^{+0.2}_{-0.3}$ & & $0.7 \pm 0.2$ & & $0.6^{+0.1}_{-0.2}$ \\
\\[-0.25cm]
\hline
\\[-0.25cm]
\xstar & & \nh & & - & & - & & - & & - & & $1.6 \pm 0.4$ \\
\\[-0.2cm]
& & $\xi$ & & - & & - & & - & & - & & $100^{+20}_{-10}$ \\
\\[-0.2cm]
& & $v$ & & - & & - & & - & & - & & $6000 \pm 4000$ \\
\\[-0.25cm]
\hline
\\[-0.25cm]
\neii\ (abs) & & $EW$ & & - & & $-3.2 \pm 0.5$ & & - & & $-2 \pm 1$ & & - \\
\\[-0.2cm]
\neiii\ (abs) & & $EW$ & & - & & $-3.2 \pm 0.6$ & & - & & $-2 \pm 1$ & & - \\
\\[-0.25cm]
\hline
\\[-0.25cm]
\pl & & $\Gamma$ & & $2.057^{+0.003}_{-0.020}$ & & $2.03^{+0.02}_{-0.03}$ & & $2.06^{+0.01}_{-0.02}$ & & $2.02 \pm 0.03$ & & $2.06^{+0.01}_{-0.03}$ \\
\\[-0.25cm]
\hline
\\[-0.25cm]
\kdblur & & \rin & & $2.2^{+0.2}_{-0.3}$ & & $1.9^{+0.6}_{-0.4}$ & & $1.56^{+0.06}_{-0.02}$ & & $1.9^{+0.6}_{-0.5}$ & & $1.7^{+0.3}_{-0.2}$ \\
\\[-0.2cm]
& & \rout & & $400$* & & $400$* & & $400$* & & $400$* & & $400$* \\
\\[-0.2cm]
& & $i$ & & $55 \pm 2$ & & $60^{+13}_{-10}$ & & $66^{+3}_{-1}$ & & $59^{+9}_{-7}$ & & $65^{+1}_{-6}$ \\
\\[-0.2cm]
& & $q$ & & $5.4^{+1.1}_{-0.2}$ & & $>4.1$ & & $7.3^{+1.1}_{-0.8}$ & & $>4.3$ & & $7^{+2}_{-1}$ \\
\\[-0.25cm]
\hline
\\[-0.25cm]
\refbhb & & $kT$ & & $0.200 \pm 0.001$ & & $0.20 \pm 0.01$ & & $0.197^{+0.001}_{-0.003}$ & & $0.20^{+0.02}_{-0.01}$ & & $0.20^{+0.01}_{-0.02}$ \\
\\[-0.2cm]
& & $n_{\rm H}$ & & $1.22^{+0.03}_{-0.04}$ & & $0.59^{+0.05}_{-0.06}$ & & $1.09^{+0.16}_{-0.08}$ & & $0.60^{+0.10}_{-0.03}$ & & $1.18^{+0.04}_{-0.30}$ \\
\\[-0.2cm]
& & $F_{\rm il}$/$F_{\rm th}$ & & $5.0^{+0.2}_{-0.1}$ & & $3.2^{+0.4}_{-0.2}$ & & $4.4 \pm 0.6$ & & $3.0^{+0.7}_{-0.3}$ & & $5.0^{+0.3}_{-2.1}$ \\
\\[-0.25cm]
\hline
\\[-0.25cm]
\oviii\ (em) & & $EW$ & & - & & - & & $60^{+10}_{-30}$ & & $15^{+9}_{-13}$ & & - \\
\\[-0.25cm]
\hline
\hline
\\[-0.25cm]
\rchi & & & & 1832/1663 & & 1721/1662 & & 1718/1662 & & 1713/1661 & & 1700/1660 \\
\\[-0.25cm]
\hline
\hline
\end{tabular}
\label{tab_refl}
\end{center}
\end{table*}

In both cases the residuals at $\sim$0.6--0.8\,\kev\ are still present. In order
to test whether these features are real, or arise due to problems with the burst
mode calibration at low energies, we also briefly analyse the 2002 \xmm\ burst
mode observations of the Crab Nebula, obtained approximately six months after
the \xte\ data presented. Data reduction was performed separately for each of
the 3 orbits following the prescription outlined in section \ref{sec_xte_red},
and the resulting spectra were combined again in a response weighted manner
using \addspec. The final spectrum obtained was modelled with an absorbed
powerlaw, again using \tbabs\ to model the Galactic absorption. We adopt the
abundances for the ISM in the direction of the crab reported by \cite{Kaastra09}
(here adopting the solar abundances of \citealt{Lodders03}, as used in that work).
The best fit column density and photon index are \nh $= 4.03 \pm 0.02 \times
10^{21}$\,\atpcm\ and $\Gamma = 2.070 \pm 0.004$, although the fit quality is
relatively poor, with \rchi = 2362/1743. \cite{epicpn_burst} also obtained a
fairly poor quality fit with a similar photon index to that obtained here. The
data/model ratio plot obtained with this model is shown in Fig. \ref{fig_xte_line}
(Panel D). It is clear that similar residuals to those seen in the \xte\ data at
$\sim$0.6--0.8\,\kev\ are present, although they do not appear to be as prominent.
This might indeed indicate the features arise at least in part due to calibration
uncertainties. Alternatively, it might indicate they originate due to
uncertainties in the absorption model used; were this to be the case the residual
features observed in \xte\ might be expected to be stronger than those in the Crab
given the higher column density obtained for \xte. Therefore, we consider it
likely that, at least to some extent, the soft residuals in \xte\ are due to
systematic uncertainties in either the \xmm\ burst mode calibration or the
absorption model. It is worth noting that residual features at soft energies have
been seen in a number of timing/burst mode observations of XRBs which may also
indicate calibration uncertainties\footnote{see http://xmm2.esac.esa.int/docs/documents/CAL-TN-0083.pdf}.
If we exclude the 0.7--1.0\,\kev\ energy range from our consideration, the simple
reflection model (Model 1) gives a good fit, with \rchi = 1672/1601.

In the interest of completeness, we also explore a number of physical explanations for
the origin of these soft features, beginning here with interpretations that do not require
additional absorption over that due to the ISM. However, we stress again our belief that
systematic uncertainties are contributing towards the residuals, so the additional
interpretations presented in this section should be treated with caution. We attempt to
model the residuals as a combination of blended \neii\ and \neiii\ absorption lines,
which are not included in \tbabs\ but frequently arise due to the ISM (\citealt{Juett06}),
and additional reflected \oviii\ emission over that expected from an accretion disc with
solar abundances, similar to that seen in the ultra-compact neutron star XRB 4U\,0614+091
(\citealt{Madej10}). In the case of 4U\,0614+091 the strong oxygen emission was explained
as arising due to accretion from an Oxygen rich white dwarf. Such a scenario is not
plausible for \xte\ so another source of oxygen rich material is required here. \refbhb\
is calculated assuming solar abundances (as given in \citealt{Morrison83}) so the extra
\oviii\ emission is modelled with the addition of a narrow Gaussian at 0.654\,\kev\ to
\refbhb\ before the relativistic effects were applied, while the Neon absorption is
modelled simply with two narrow Gaussian absorption lines, required to contribute equally
for simplicity (the work of \citealt{Juett06} suggests this is not unreasonable). This
combination provides a good representation of the data, with an improvement of 119 for 2
extra degree of freedom, and the parameters obtained are given in Table \ref{tab_refl} as
Model 4.

On investigation, we find that either the combined Neon absorption lines or the additional
Oxygen emission can resolve the residuals individually, providing improvements of
$\Delta$\chisq = 111 and 114 (for one additional degree of freedom) respectively. For
comparison, the parameters obtained with these interpretations are also given in Table
\ref{tab_refl} as Model 2 and Model 3 respectively. However, in the former case the Neon
equivalent widths are $EW$ = -$55\pm10$\,m\AA\ (for each line), much greater than those
obtained using the 2001 \chandra\ MEG spectra of \xte\ ($\sim$ -2--10\,m\AA;
\citealt{Miller04}). In the latter, the disc inclination is strongly constrained to be
high ($65 < i < 69$\deg), inconsistent with that obtained with a similar
interpretation in \cite{Miller09}, although we note that a different reflection code is
used in this work, and the inclination obtained here is consistent with the lower limit
of $50\pm3$\deg\ obtained from the optical light curve by \cite{Orosz04}. We consider a
combination of these features more plausible, and with this interpretation neither the
inclination or the Neon equivalent widths are as tightly constrained, due to the
degeneracy between the inclination and the emissivity index when modelling the line
profile, and also the degeneracy between the individual interpretations. The Neon
equivalent widths may be made consistent with those presented in \cite{Miller04} by
shifting more emphasis onto the Oxygen emission. However, we note that doing so tightens
the requirement for a high inclination. It seems that in assigning a combination of
these physical origins for the soft residuals this model can either be consistent with
the inclination obtained in \cite{Miller09} or the Neon absorption presented in
\cite{Miller04}, but it cannot be consistent with both. 

It is also possible that the soft residuals may indicate absorption by ionised
material associated with \xte\ rather than the ISM, most likely due to an
outflowing disc wind. To test this we generated a grid of ionised absorption
models using \xstar v2.2.1, adopting solar abundances for all the elements and
a turbulent velocity of 100\,\kmps. The free parameters of this model are the
column density, the redshift, which may in turn be used to infer the outflow
velocity, and the ionisation parameter of the absorbing medium. Including this
additional component in the basic disc reflection interpretation constructed
previously, the model also provides a good representation of the data, with an
improvement of $\Delta$\chisq = 126 for 3 additional degrees of freedom. The
key parameters for this interpretation are also given in Table \ref{tab_refl},
as Model 5.

\begin{figure}
\begin{center}
\rotatebox{0}{
{\includegraphics[width=230pt]{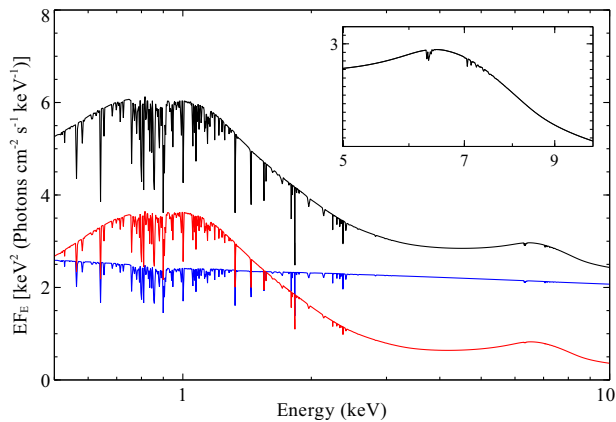}}
}
\end{center}
\caption{The model obtained treating the soft residuals as evidence for an ionised
disc wind (see \S \ref{sec_abs} and Model 5 in Table \ref{tab_refl}) through use
of an \xstar\ absorption grid. The total model, corrected for Galactic absorption,
is shown in black, the red shows the contribution of the reflection component
(including the thermal emission from the disc) and the blue shows the powerlaw
continuum. \textit{Inset:} the same model zoomed in on the 5--10\,\kev\ energy range.
The ionised absorber does not contribute any strong absorption features at these
energies.}
\label{fig_abs1}
\end{figure}

In this scenario, the residuals represent a complex blend of various narrow
absorption features. The best fit value obtained for the redshift is
\textit{positive}, which suggests that if the material is associated with \xte\
it is \textit{infalling}, rather than outflowing, with a velocity of $6000 \pm
4000$\,\kmps\ (although this is not particularly well constrained due to the
blend of multiple features contributing). Given that the work of \cite{Orosz04}
strongly suggests \xte\ is a low mass X-ray binary (LMXB), and hence mass
transfer should occur through Roche-lobe overflow rather than via stellar winds,
the presence of such infalling material might be considered unusual, especially
given the high velocity, so it is not clear this is the correct interpretation,
although it remains an intriguing possibility. Such material could be the remnant
of a failed outflow, which is now falling back onto the black hole. However,
even if the soft residuals are due to absorption by partially ionised matter,
Fig. \ref{fig_abs1} shows that this material does not contribute any significant
absorption features in the $\sim$3--10\,\kev\ energy range.

In summary, a number of astrophysical origins for the soft residuals provide
statistically acceptable fits to the data. However, given the physical issues
each of these raise (requiring unreasonably high neon absorption in the ISM
and/or an unexplained over abundance of oxygen, or the presence of rapidly
infalling material) and the comparison with the burst mode data from the Crab
nebula, we strongly suggest these features are systematic rather than astrophysical.
The roughly consistent inner radius obtained with each interpretation demonstrates
that they are not critical to our determination of this key parameter.

\subsubsection{Ionised Absorption in the Iron K Band}
\label{sec_abs}

We now focus on the 5--10\,\kev\ range in order to determine whether absorption
from highly ionised species could modify the obtained line profile. The only possible
absorption feature in the data is at $\sim$7.1\,\kev. Assuming an identification with
hydrogenic iron, this material would be outflowing at $6000 \pm 2000$\,\kmps, much
higher than the typical velocities seen for XRB disc winds, which are less than
$\sim$1000\,\kmps\ (see \eg \citealt{Lee02}, \citealt{Ueda04}, \citealt{Miller06a},
\citealt{Neilsen09}), but not as high as the relativistic wind described in
\cite{Done06} (see below). However, the detection is fairly tentative (including an
unresolved Gaussian absorption line gives an improvement of $\Delta$\chisq $\sim$\,8
for 2 extra degrees of freedom) and if real this is a weak feature
($EW$\,$\sim$\,-10\,eV); its inclusion in any of the presented models does not alter
the inner radius obtained. If this absorption line is real, and ionised absorption is
the correct interpretation for the soft residuals, the significant difference in
their velocities suggests these features cannot be associated with the same absorber.

\begin{figure}
\begin{center}
\rotatebox{0}{
{\includegraphics[width=230pt]{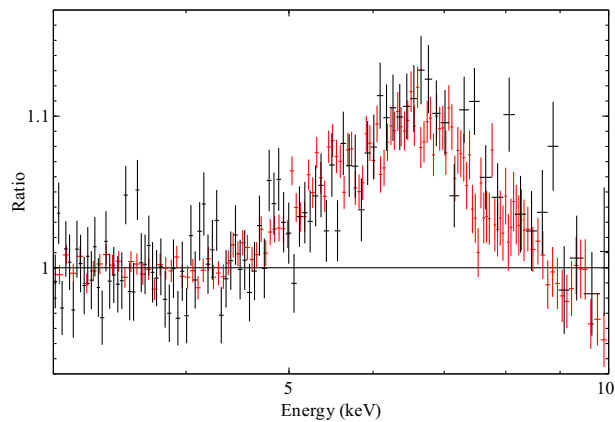}}
}
\end{center}
\caption{iron line profiles for the \xmm\ (black) and \bepposax (red) observations
of \xte, obtained on 2001 September 13th and 11th respectively. The two
observations may be fit with similar continuum models, resulting in practically
identical line profiles.}
\label{fig_1650_sax}
\end{figure}

Based on the \bepposax observation of \xte\ taken about a day before the \xmm\
observation analysed here, \cite{Done06} argue that a strong, highly
blueshifted absorption line at $\sim$7.6\,\kev\ may be present due to ionised
iron in a relativistically outflowing disc wind, and that failing to account
for this line can lead to an artificially low inner disc radius. The line can
instead be consistent with a truncated disc of inner radius $\sim$10--20\,\rg,
although reflection of the Comptonised continuum by this disc, producing a
fairly broad iron line, is still required. We extracted the \bepposax MECS spectrum in
question (covering 2.5--10.0\,\kev), and phenomenologically modelled the
2.5--4.0 and 8.0--10.0\,\kev\ data with the basic BHB components as described
in \S \ref{sec_xte_cont}; the continuum model obtained for this observation is
very similar to that for the \xmm\ data. A broad excess emission feature is
again clearly apparent when the 4.0--8.0\,\kev\ data are included. In Fig.
\ref{fig_1650_sax} we show both the \bepposax and \xmm\ line profiles; the two
are found to be practically identical.

We then verified that the MECS data, covering 2.5--10.0\,\kev\ can indeed be
consistent with a truncated disc when ionised absorption from a relativistic
outflow is included, making use again of \refbhb\ and the \xstar\ grid
described above. Throughout this analysis a lower limit of 45\deg\ was imposed
on the inclination, similar to the lower limit found by \cite{Orosz04}, and
the same interstellar absorption as obtained with the full \xmm\ data was
applied. We obtained an inner disc radius of $16^{+4}_{-3}$\,\rg\ for the MECS
data, with absorber properties $N_H = 2.2^{+0.6}_{-0.4} \times 10^{22}$\,\atpcm,
$\xi = 240 \pm 60$\,\ergcmps\ and $v_{\rm out} = 0.138 \pm 0.007$\,$c$. These
values are similar to those presented in \cite{Done06}.

\begin{figure}
\begin{center}
\rotatebox{0}{
{\includegraphics[width=230pt]{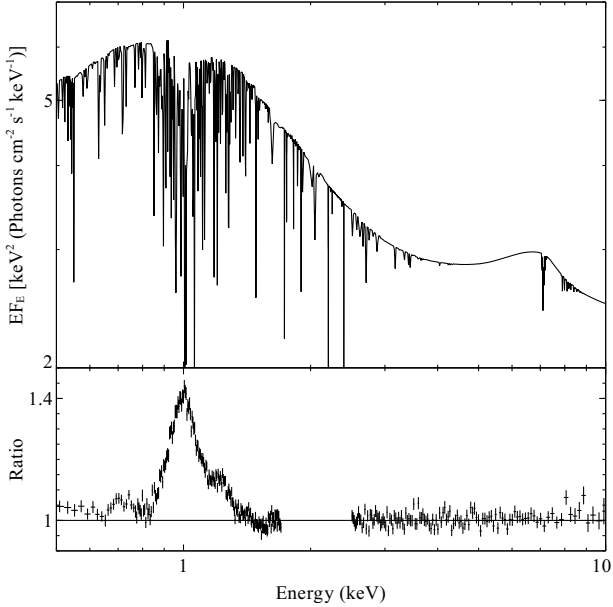}}
}
\end{center}
\caption{\textit{Top panel}: The model obtained for \xte\ considering only the
\xmm\ data above 2.5\,\kev\ and including a strong relativistic outflow, similar
to that proposed in \citet{Done06}, in order to allow the line profile to be
consistent with a truncated disc. This outflow would also imprint strong absorption
features below 2.5 \,\kev. \textit{Bottom panel}: Data/model ratio plot of the
broad band \xmm\ spectrum to the model shown in the top panel, with the
0.5--1.7\,\kev\ data included. The low energy absorption predicted with the
inclusion of a strong, relativistic outflow is clearly not present in the data.}
\label{fig_abs2}
\end{figure}

On investigation, even though there is no strong visual hint of any strong absorption
in the high energy data, we also found that the 2.5--10.0\,\kev\ \xmm\ spectrum can
be consistent with a similar interpretation (\rin = $17^{+92}_{-9}$\,\rg), with \nh
= $10^{+8}_{-5} \times 10^{21}$\,\atpcm, $\xi = 140^{+70}_{-120}$\,\ergcmps\ and
$v_{\rm out} = 8.8^{+0.6}_{-1.0} \times 10^{-2}$\,$c$. However, as shown in Fig.
\ref{fig_abs2} (top panel), this absorber predicts a variety of additional absorption
features below 2.5\,\kev. Including again the 0.5--1.7\,\kev\ \xmm\ data, it is clear
these absorption features are not present in the data, with the model providing a
very poor fit over this energy range (\rchi = 15477/1665) as demonstrated by Fig.
\ref{fig_abs2} (bottom panel). Finally we attempted to model the full \xmm\ spectrum
with this interpretation, resulting in the inner radius settling again at
$\sim$2\,\rg\ and the contribution of the absorber being minimised. Although the
interpretation invoking a combination of a truncated disc and a relativistic outflow
remains viable if only the 2.5--10.0\,\kev\ data are considered, the additional data
below 2.5\,\kev\ provided by \xmm\ excludes the presence of absorption strong enough
to affect the observed width of the iron line. Therefore, we consider the inner
radius of $\sim$2\,\rg\ consistently obtained to be robust to the effects of
absorption.

\subsubsection{Instrumental Effects}
\label{sec_inst}

As stated previously, the residual features seen in the $\sim$1.7--2.5\,\kev\
range are most likely due to calibration uncertainties associated with the silicon K
and gold M instrumental edges. These features are most prominent for sources with high
countrates, such as \xte, and it has been suggested that this may be (at least partly)
due to residual charge transfer inefficiency\footnote{see http://xmm2.esac.esa.int/docs/documents/CAL-TN-0083.pdf}
(hereafter CTI). This is a process that occurs during the readout of CCD detectors. As
charge is transferred from one pixel to another, some of the electrons may be
lost into `charge traps' randomly distributed throughout the silicon lattice. This
results in the measured energy of the incident photon being underpredicted. More
recently, evidence has been found that suggests the residuals may instead be caused by
X-ray loading\footnote{http://xmm2.esac.esa.int/docs/documents/CAL-TN-0050-1-0.ps}
(hereafter XRL). This effect is caused by X-ray photons contaminating the offset map
produced prior to each observation to determine the energy zero-point for each pixel,
and hence occurs preferentially for bright/hard sources. Whatever the physical cause,
the current corrections applied in the standard \epicpn\ data reduction procedure do
not adequately describe the instrumental features for high count rates.

\begin{figure}
\begin{center}
\rotatebox{0}{
{\includegraphics[width=230pt]{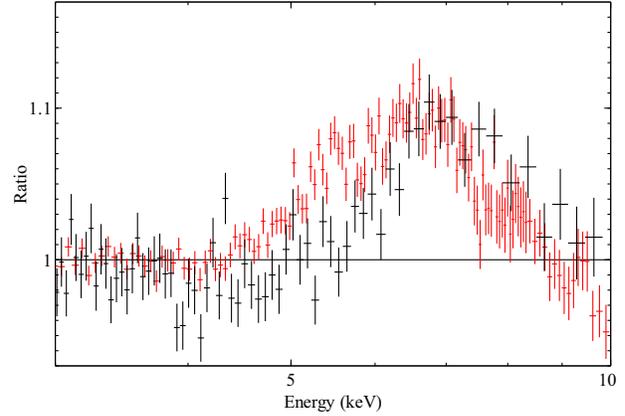}}
}
\end{center}
\caption{Iron line profiles for the \xmm\ (black) and \bepposax (red) observations
of \xte, as in Fig. \ref{fig_1650_sax}, but here the \xmm\ data has been modified
with the \epfast\ tool. This modification results in clearly discrepant line profiles.}
\label{fig_epfast}
\end{figure}

\begin{figure*}
\begin{center}
\rotatebox{0}{
{\includegraphics[width=220pt]{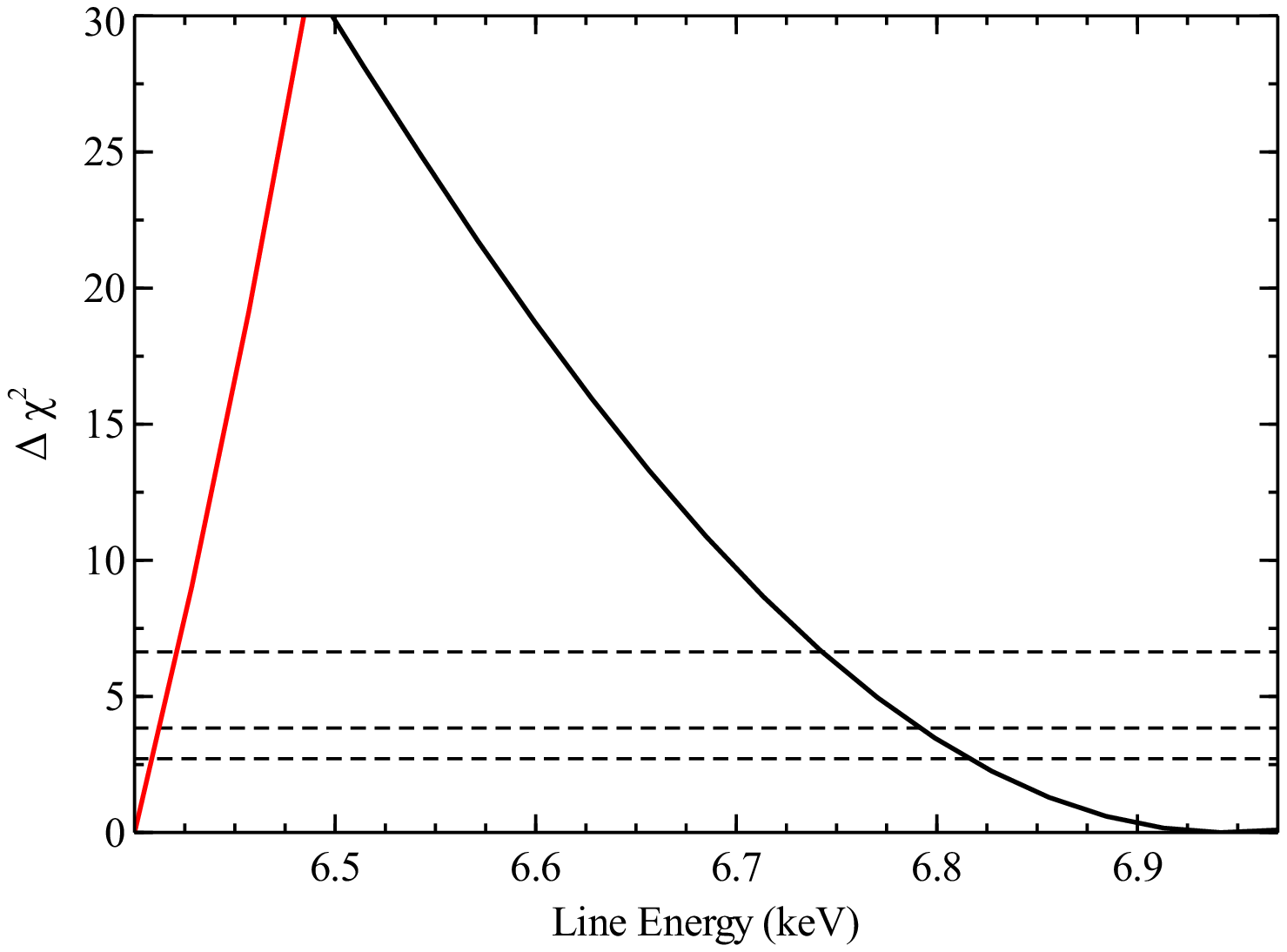}}
}
\hspace{1cm}
\rotatebox{0}{
{\includegraphics[width=220pt]{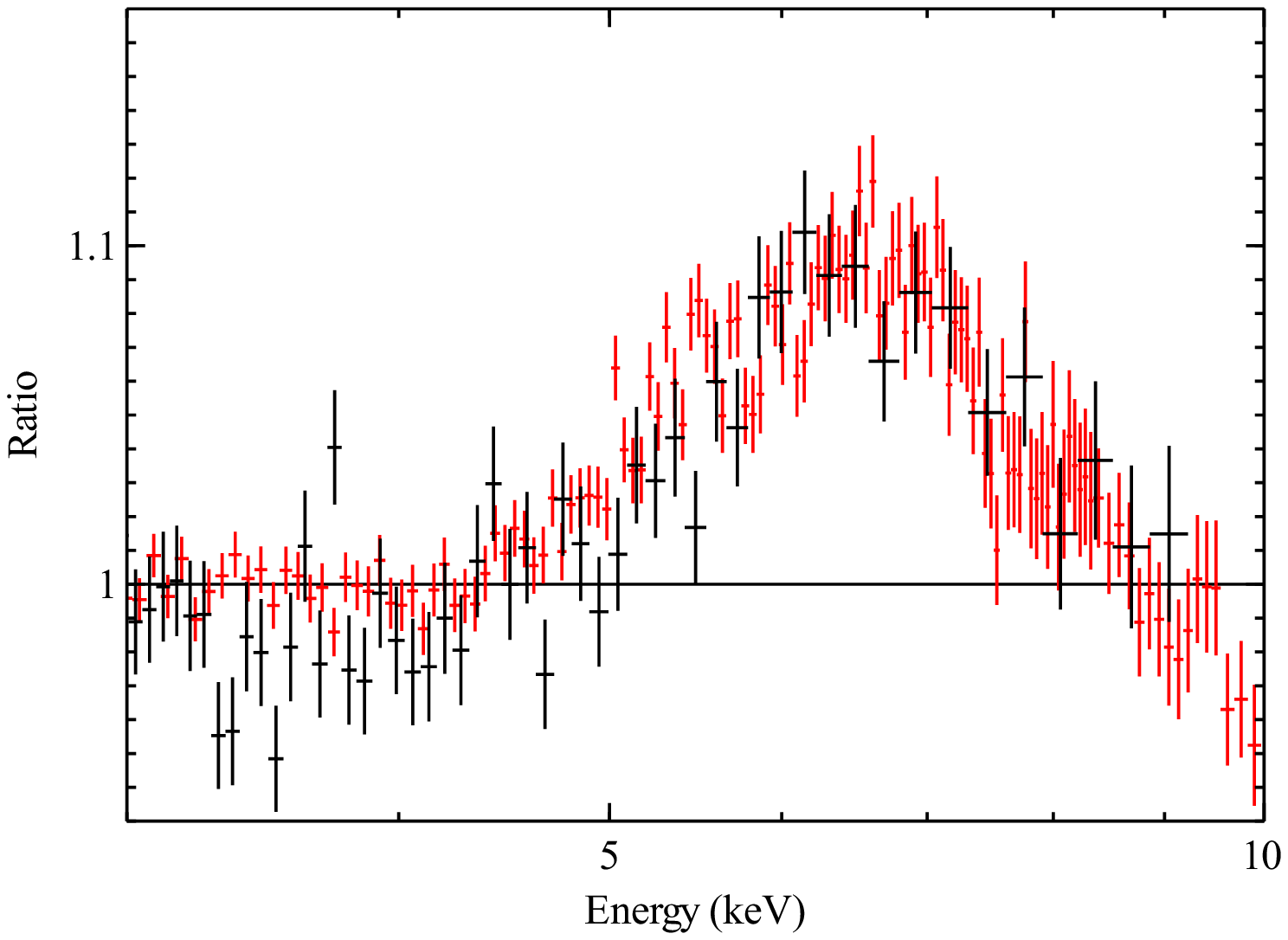}}
}
\end{center}
\caption{\textit{Left panel}: \chisq confidence contours for the line energies obtained
with the \epfast-modified \xmm\ (black) and \bepposax (red) datasets when modelling the
iron emission with a simple Gaussian line profile. Horizontal dased lines show the 90,
95 and 99 per cent confidence intervals for one parameter of interest. The line energies
obtained with the two datasets are clearly not in agreement. \textit{Right panel}:
Comparison of the \xmm\ \epfast-modified (black) and \bepposax (red) line profiles, with
the energy scale of the former shifted by $\sim$0.55\,\kev, as suggested by the left panel.
Including this energy shift returns the profiles obtained to excellent agreement.}
\label{fig_lineE}
\end{figure*}

The \epfast\ tool marks a serious effort to improve the performance of \epicpn\
fast-readout modes (Timing and Burst) by accounting for these remaining instrumental
effects. In our reduction and analysis, we have found that \epfast\ produces spectra
that are better described by standard spectral responses at low energies, specifically
in the 1.5--2.5\,keV range where it corrects for the apparent shift of the instrumental
edges very well. However, \epfast\ is likely unsuited to higher energies at present,
because it applies an energy independent correction ($\Delta E/E$ = constant). Studies
show that CTI depends on energy as $(\Delta E/E)_{\rm CTI} \propto E^{-\alpha}$, with
$\alpha \simeq 0.5$, meaning that the relative impact of CTI is 2 times greater at
2\,keV than at 6--7\,keV (see \eg \citealt{Gendreau93cti}, \citealt{Townsley02cti},
\citealt{Grant04cti}, \citealt{Prigozhin04cti}, \citealt{Nakajima08cti}). Indeed, the
CTI correction included in the standard reduction procedures is an energy dependent
correction, but any remaining CTI will also be energy dependent. If instead the
residuals are due to XRL, the correction applied by \epfast\ is an even worse
approximation. To zeroth order, XRL causes a constant offset in the energy scale, i.e.
$\Delta E$ = constant, therefore in this case, again writing $(\Delta E/E)_{\rm XRL}
\propto E^{-\alpha}$, we have $\alpha \simeq 1$, hence the relative impact of XRL at
6\,\kev, in comparison to at 2\,\kev, is even less than for CTI. As the level of
correction is primarily calibrated at $\sim$2\,keV, especially for burst mode,
\epfast\ must over-correct spectra at high energies.

This is supported by the striking difference between the \bepposax and
\epfast-modified \xmm\ spectra of \xte\ in the iron K band (see Fig. \ref{fig_epfast}).
In contrast, as demonstrated by Fig. \ref{fig_1650_sax}, the unmodified \xmm\ spectrum
is remarkably similar to the \bepposax data. As the \xmm\ and \bepposax observations
in question are only separated by a day, given the long term evolution of \xte\ during
this outburst presented by \cite{Done06}, we expect any astrophysical changes displayed
by \xte\ between these two observations to be minimal. The disagreement between the
line profiles in the \epfast-modified \epicpn\ and the \bepposax spectrum, as well as the
agreement between the profiles in the latter and the unmodified \epicpn\ spectrum, suggest
that the unmodified data may represent a better approximation of the true energy scale
in the $\sim$6\,\kev\ energy range for count rates as high as measured in \xte. In an
attempt to quatify the comparisons presented in Figs. \ref{fig_1650_sax} and
\ref{fig_epfast} in a simple manner, we return to the Gaussian line profile considered
initially in section \ref{sec_xte_line}, and applied this model to both the
\epfast-modified and \bepposax data. The results obtained for these data, as well as
those from the unmodified \xmm\ data obtained previously, are presented in Table
\ref{tab_gauss}. As before, the line energies were constrained to be in the range
6.4--6.97\,\kev.

The line width obtained with the \epfast-modified data is marginally smaller than that
obtained with the unmodified data. Although this must be considered a systematic change,
as it arises due to a redistribution of photon energies, the difference is small in
comparison with the statistical uncertainty on each; the widths of the two profiles are
fully consistent within their 90 per cent confidence limits, and both values are extremely
broad. Therefore, we conclude that in this case instrumental effects do not appear to be
causing a statistically significant broadening of the line profile. However, when
considering the line energies obtained, only the unmodified \xmm\ data is in agreement
with \bepposax. The disagreement between the \epfast-modified \xmm\ and the \bepposax data
is further emphasised in Fig. \ref{fig_lineE} (\textit{left panel}), which shows the line
energy confidence contours for the two datasets. The main effect of \epfast\ in this case
is to shift the line profile to higher energies, without significantly modifying its width.
To further demonstrate this, in Fig. \ref{fig_lineE} (\textit{right panel}) we again
overlay the \epfast-modified and \bepposax line profiles, but now with the former shifted
down in energy by $\sim$0.55\,\kev\ in order to realign the line centroids; the profiles
are returned to excellent agreement. Indeed, modelling the line with a \laor2\ profile at
7.5\,\kev\ (again shifted by $\sim$0.55\,\kev\ in comparison to the unmodified case) gives
practically identical results to those presented in Table \ref{tab_laor}: \rin =
$1.37^{+0.06}_{-0.02}$\,\rg, $q_{\rm inner} > 8.1$ and $i = 70\pm1$\deg. Although a shift
of 0.55\,\kev\ represents an energy difference of $\sim$8 per cent and \epfast\ only
formally applies a gain shift of $\sim$3 per cent in this case, in appendix A we demonstrate
that the use of \epfast\ naturally leads to a much larger energy shift when modelling a
broad feature at 6.4\,\kev, fully consistent with that obtained, as the application of this
tool modifies the high energy continuum, which can be difficult to determine underneath such
a broad line.

Based on this comparison and the theoretical considerations of the relevant instrumental
effects discussed earlier, we conclude that the unmodified data do provide the best
approximation of the true energy scale, while the modified data probably yield the best
description of the line width. Furthermore, the good agreement of the \bepposax
and unmodified \xmm\ spectra suggests that some process other than CTI or XRL is the primary
cause of the instrumental residuals at $\sim$2~keV, the overall effect of this process at
high energies ($\sim$\,6\,\kev) is small. We note briefly that if either CTI or XRL is the
dominant cause of the residuals at $\sim$2\,\kev, as appears likely, these effects could
also be contributing to the soft residuals observed below 1\,\kev, given that the relative
strength of both of these processes is higher at soft energies. These features are of
little consequence for the main aim of this paper, but we again urge that the possible
physical origins presented for the soft residuals be treated with caution.

\begin{table}
  \caption{Comparison of the results obtained for the unmodified and \epfast-modified
\xmm\ calibrations and the contemporaneous \bepposax observation, when modelling the
iron emission with a simple Gaussian line profile. The line widths are consistent in
all cases, but only the unmodified \xmm\ calibration provides a statistical agreement
with the \bepposax data for the line energy (see also Fig. \ref{fig_lineE}).}
\begin{center}
\begin{tabular}{c c c}
\hline
\hline
\\[-0.25cm]
Dataset & $E_{\rm Gauss}$ & $\sigma$ \\
\\[-0.2cm]
& (keV) & (keV) \\
\\[-0.25cm]
\hline
\\[-0.25cm]
\bepposax (11th September) & $< 6.41$ & $1.09 \pm 0.1$ \\
\\[-0.2cm]
\xmm\ (unmodified) & $< 6.50$ & $1.17^{+0.24}_{-0.21}$ \\
\\[-0.2cm]
\xmm\ (\epfast-modified) & $> 6.81$ & $1.07 \pm 0.25$ \\
\\[-0.25cm]
\hline
\hline
\end{tabular}
\label{tab_gauss}
\end{center}
\end{table}

In the interest of completeness, and for further comparison, we also present the results
obtained with the basic reflection interpretation (Model 1) applied to the \epfast-modified
data, including the 1.7--2.5\,\kev\ energy range. An excellent fit is obtained, with \rchi
= 1843/1840. The values obtained for various key parameters are quoted in table
\ref{tab_epfast}, and we also list again the values obtained with the unmodified data for
ease of comparison. We only consider the basic reflection interpretation here given the
likely systematic origin of the soft residuals. In any case, with the \epfast-modified data
it is possible to resolve these residuals merely as absorption by the Galactic ISM, although
some unusual abundances are required, potentially suggesting that the energy scale below
$\sim$2\,\kev\ may also be incorrect. Our discussion henceforth is limited to the parameters
that dictate the relativistic blurring, and hence determine the line profile: the main focus
of this work.

\begin{table}
  \caption{Comparison of the parameter values obtained for the basic reflection
interpretation with both the unmodified and the \epfast-modified \xmm\ data. As with
Table \ref{tab_refl}, column densities are given in $10^{21}~$\atpcm, abundances are
quoted relative to the solar abundances, radii in gravitational radii, inclinations
in degrees, temperatures in \kev\ and hydrogen number densities are given in $10^{20}$
atom cm$^{-3}$.}
\begin{center}
\begin{tabular}{c p{0.2cm} c p{0.2cm} c}
\hline
\hline
\\[-0.25cm]
Parameter & & Unmodified & & \epfast-modified \\
\\[-0.25cm]
\hline
\\[-0.25cm]
\nh & & $7.17^{+0.08}_{-0.09}$ & & $6.7 \pm 0.1$ \\
\\[-0.2cm]
$A_{\rm O}$ & & $1.06^{+0.05}_{-0.02}$ & & $1.36^{+0.06}_{-0.04}$ \\
\\[-0.2cm]
$A_{\rm Ne}$ & & $1.36^{+0.06}_{-0.03}$ & & $1.7 \pm 0.1$ \\
\\[-0.2cm]
$A_{\rm Fe}$ & & $0.95^{+0.04}_{-0.07}$ & & $0.52^{+0.08}_{-0.04}$ \\
\\[-0.2cm]
$\Gamma$ & & $2.057^{+0.003}_{-0.020}$ & & $1.99^{+0.01}_{-0.02}$ \\
\\[-0.2cm]
\rin & & $2.2^{+0.2}_{-0.3}$ & & $3 \pm 1$ \\
\\[-0.25cm]
$i$ & & $55 \pm 2$ & & $85^{+2}_{-11}$ \\
\\[-0.25cm]
$q$ & & $5.4^{+1.1}_{-0.2}$ & & $2.0^{+0.2}_{-0.3}$ \\
\\[-0.25cm]
$kT$ & & $0.200 \pm 0.001$ & & $0.200 \pm 0.002$ \\
\\[-0.2cm]
$n_{\rm H}$ & & $1.22^{+0.03}_{-0.04}$ & & $1.34^{+0.07}_{-0.05}$ \\
\\[-0.2cm]
$F_{\rm il}/F_{\rm th}$ & & $5.0^{+0.2}_{-0.1}$ & & $5.0^{+0.6}_{-0.1}$ \\
\\[-0.2cm]
\hline
\hline
\end{tabular}
\label{tab_epfast}
\end{center}
\end{table}

First of all, we note the consistency of the values obtained for the inner radius; the radius
obtained with the \epfast-modified data is marginally larger, but the statistical agreement
with the value obtained with the unmodified data is excellent. Given the similarly excellent
agreement of the line widths when using the simple Gaussian model, this is not surprising.
There are, however, two significant differences that require highlighting. The inclination
is significantly higher and the index of the emissivity profile significantly lower when
\epfast\ is applied. This is a direct consequence of the line profile being shifted to
higher energies by \epfast. As we are using a physically self-consistent reflection model,
the intrinsic energy of the iron emission is hardwired into the model, and only the blurring
kernel (\kdblur) is able to change the energy at which the emission is observed. In order
to shift the line to higher energies, the blurring kernel must both enhance the blue wing
(increase the inclination) and attempt to suppress some of the red wing (decrease the
emissivity index) of the line profile. The best fit inclination obtained using the \epfast\
tool is unphysically high, implying that we are observing the source practically through the
plane of the disc, although the 90 per cent confidence limit does extend to slightly less
extreme inclinations. Therefore, the incorrect energy scale of the \epfast-modified data at
high energies can have a significant impact on key physical parameters when using physical
rather than phenomonological models. We stress again that these changes are not physical,
and are purely a result of the line profile being incorrectly shifted to higher energies
when the \epfast\ tool is applied.

To summarise briefly, we have investigated in detail the application of the \epfast\ tool,
which was designed to correct the instrumental features observed in \epicpn\ data around
$\sim$2\,\kev\ at high count rates. The correction applied by \epfast\ does not -- at the
time of writing -- have the correct energy dependence for either CTI or XRL, the two likely
physical causes of these features, and although this does not lead to any significant
changes in the width of the observed line profile in this case -- which is ultimately the
main focus of this work -- it does result in an incorrect energy scale at high energies.
As demonstrated through comparison with a contemporaneous \bepposax observation and our
simulations in Appendix A, \epfast\ incorrectly shifts the line profile to higher energies.
This energy shift can have important implications for key parameters when using physically
self-consistent reflection models in which the line energy is not a free parameter. We will
proceed henceforth with the use of the unmodified \xmm\ data.

\subsubsection{Alternative Scenarios}
\label{sec_alt}

Until now we have been considering the case in which the line arises due to
reflection close to the black hole and is broadened by the relativistic effects
present in such a region. This model therefore requires that the accretion disc,
which in this scenario is the reflector, extends in (or close to) to the ISCO.
However, as stated previously, there is some debate over whether this is the
case at low accretion rates (specifically in the low/hard state). Having
demonstrated that the line is not artificially broadened by instrumental effects,
we now investigate whether the line can be primarily broadened by some other
physical mechanism. The main alternative is that broadening via electron scattering
(Compton broadening) dominates. This process gives a line width per scattering,
$\sigma$, of:

\vspace{-0.35cm}
\begin{equation}
\frac{\sigma}{E} \simeq \left ( \frac{2kT_{\rm e}}{m_{\rm e}c^{2}} \right )^{0.5} 
\label{eqn_compt1}
\end{equation}

\noindent{where $E$ is the energy of the line and $T_{\rm e}$ is the electron
temperature (\citealt{Pozdnyakov83}). In addition to broadening discrete features,
Compton scattering also amplifies the energy centroid by:}

\vspace{-0.35cm}
\begin{equation}
\frac{E_{\rm obs}}{E_{\rm int}} = 1 + 4 \left ( \frac{kT_{\rm e}}{m_{\rm e}c^{2}} \right ) + 16 \left ( \frac{kT_{\rm e}}{m_{\rm e}c^{2}} \right )^{2} + ...
\label{eqn_compt2}
\end{equation}

\begin{figure}
\begin{center}
\rotatebox{270}{
{\includegraphics[width=155pt]{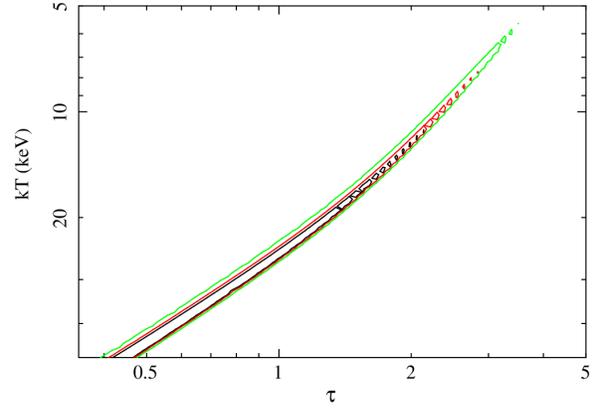}}
}
\end{center}
\caption{\chisq\ confidence contours for the optical depth ($\tau$) and electron
temperature ($kT_{e}$) obtained for the Comptonising medium with \comptt\ for the
\xmm\ observation of \xte. The contours shown represent the 90 (black), 95 (red)
and 99 (green) per cent confidence contours for two parameters of interest.}
\label{fig_compt}
\end{figure}

\begin{figure*}
\begin{center}
\rotatebox{0}{
{\includegraphics[width=470pt]{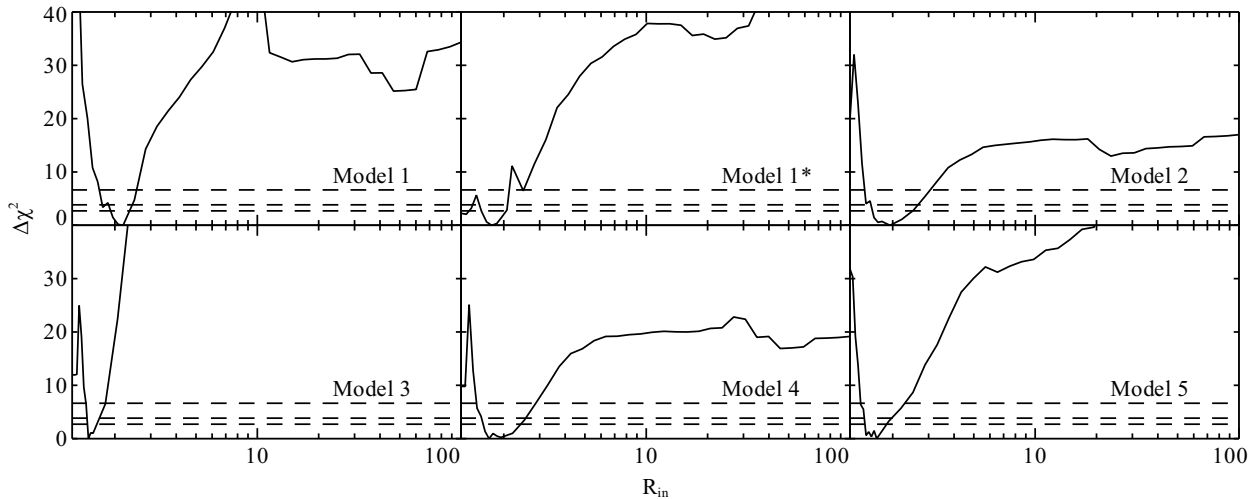}}
}
\end{center}
\caption{\chisq\ confidence contours for the inner radius for each model presented
in Table \ref{tab_refl}. The horizontal dashed lines represent $\Delta$\chisq\
equivalent to the 90, 95 and 99 per cent confidence intervals. Small values for the
inner radius are significantly favoured in each case.}
\label{fig_Rin}
\end{figure*}

\noindent{where $E_{\rm obs}$ and $E_{\rm int}$ are the observed and intrinsic
line energies respectively. Considering the likely geometry and processes
occurring in BHBs, Compton broadening could potentially occur in three separate
media: the disc itself, the corona, or some outflowing disc wind. We will consider
these three cases individually.}

First we consider the scenario in which the breadth of the line is primarily due to
Compton broadening in the corona. Modelling the line with a Gaussian profile, we
obtained a line energy of $\sim$6.4\,\kev\ and a width of $\sigma \sim1.1$\,\kev\
(see sections \ref{sec_xte_line} and \ref{sec_inst}). As there is no obvious
narrow component to the line profile, invoking Compton broadening as the origin of
its width requires that the optical depth of the corona $\tau \gtrsim$\,1, \ie the
vast majority of the line photons must experience at least one scattering. Here we
consider the simplest case, in which $\tau \sim 1$ and the line is broadened by a
single scattering process. Substituting the values obtained from the Gaussian profile
into equation \ref{eqn_compt1} we find that, in order for this to be the case,
$kT_{\rm e} \simeq$\,6\,\kev. If we wish to invoke multiple scatterings, \ie
increasing the optical depth, in order to obtain the same line width, $T_{\rm e}$
would have to be lower still. However, we remind the reader that Comptonisation of
the line in the corona reduces its observed equivalent width by a factor of $e^{\tau}$
(see \citealt{Petrucci01}), so the case with $\tau \sim$\,1 is the most reasonable.

A corona with $\tau$\,$\sim$\,1 and $kT_{\rm e}$\,$\sim$\,6\,\kev\ would
produce a curved continuum in the \xmm\ bandpass. However, in section \ref{sec_compt}
we demonstrated that the continuum is powerlaw-like, with minimal curvature. To
further demonstrate that such a continuum is not observed, Fig. \ref{fig_compt}
shows the confidence contours for $\tau$ and $kT_{\rm e}$ when modelling the high energy
spectrum with a combination of \comptt\ and a Gaussian emission line. A combination
of $\tau \sim$\,1 and $kT_{\rm e} \sim$\,6\,\kev\ is clearly excluded. In addition,
considering equation \ref{eqn_compt2}, Comptonisation of the line in such a corona
should shift the line centroid to $\sim$6.75\,\kev\ (assuming an initial energy of
6.4\,\kev). However, the line energy is constrained to be below 6.5\,\kev. Therefore
Compton broadening in the corona does not produce a self consistent interpretation
for both the line profile and the observed continuum. It is worth noting that
\cite{Hiemstra11} present a similar consideration for the high energy spectrum of
XTE\,1652-453, and also conclude that the line profile in that source cannot be
explained by Compton broadening in the corona.

Second we consider electron scattering in the disc itself. As previously noted,
this can be an important process in BHBs given their high disc temperatures. Indeed,
we stress again that the reflection model used here, \refbhb, has been calculated
specifically for use with spectra from XRBs, and self-consistently includes this
process at an appropriate level for such sources. In order to demonstrate that
relativistic effects are preferred over Compton processes in the disc, in Fig.
\ref{fig_Rin} we show the \chisq confidence contours for the inner radius up to
100\,\rg\ for each model presented in Table \ref{tab_refl}. In addition, we also
show the case of the basic reflection model but with the $\sim$0.7--1.0\,\kev\ data
excluded, referred to as Model 1*, to demonstrate that the presence of the soft
residuals has little effect on the value obtained for the inner radius with this
model. Increasing the inner radius has the effect of shifting the emphasis from a
combination of Doppler and gravitational broadening to a combination of Doppler and
Compton broadening. Clearly a small inner radius, and therefore the former scenario,
is favoured at greater than 99 per cent confidence in each case. Furthermore,
\cite{Reis09spin} demonstrate that solutions invoking highly ionised reflection at
large distances from BHBs are not physically consistent with the expectation of a
standard thin disc.

Finally, we also briefly consider the possibility that the line is not produced
by a reflection process in the disc, but is instead produced via reprocessing of
the X-rays from the central source in an outflow, then broadened by electron
scattering as it escapes from the outflowing medium (see \eg
\citealt{Titarchuk09}). Such a model may statistically reproduce the line
profiles observed in a variety of neutron star and black hole XRBs. However,
as shown by \cite{Cackett10}, when considered in detail there are a number of
issues with this `windline' interpretation. We refer the reader to that work
for the full discussion, and merely summarize the main points here. First and
foremost, in order to reproduce the line profiles observed the outflows must
be optically thick. However, broad iron lines are observed in XRBs across a
broad range in inclination, hence if such lines are produced in a disc wind
these outflows must cover a large solid angle. This is in contrast with both
theoretical simulations of disc winds and observational evidence for absorption
by such outflows; see section \ref{sec_disc} for a full consideration of the
(lack of) connection between broad emission lines and disc winds.

In addition, the mass outflow rates obtained in \citealt{Titarchuk09} for the
BHB GX\,339-4 and the Neutron star (NS) XRB Ser\,X-1 are both comparable with
the Eddington mass accretion rate. However, in the observations analysed in
that work both sources are known to be accreting comfortably below the
Eddington limit, so the mass outflow rate required by the windline
interpretation exceeds the mass inflow rate due to accretion\footnote{Although
the authors of that work make a claim to the contrary, they incorrectly compare
the mass outflow rate to the inflow rate for the high/soft state, despite the
wind parameters being obtained for low/hard state observations, in which the
accretion rate is much lower.}. Furthermore, broad iron lines are observed in
NS XRBs across at least two orders of magnitude in luminosity, therefore the
high mass outflow rate would need to be independent of the accretion rate. This
requirement is further supported by the lack of significant evolution in the
line profile with accretion rate displayed by \xte\ (see section
\ref{sec_line_evol}) and XTE\,J1752-223 (\citealt{Reis1752}). Finally, broad
lines and high frequency quasi-periodic oscillations (HFQPOs) are observed
simultaneously in NS XRBs (see \eg \citealt{Cackett09a}). For any viable HFQPO
formation mechanism, their presence dictates that the inner disc must be directly
observed, rather than obscured by an optically thick outflow. In addition,
\cite{Hiemstra11} argue that it is difficult to reconcile a Compton thick outflow
with a Compton thin corona. Taking all this into consideration, we conclude that
the windline interpretation is not a viable solution for the presence of broad
iron lines in XRBs.

\subsubsection{Line Profile Evolution}
\label{sec_line_evol}

\begin{figure}
\begin{center}
\rotatebox{0}{
{\includegraphics[width=230pt]{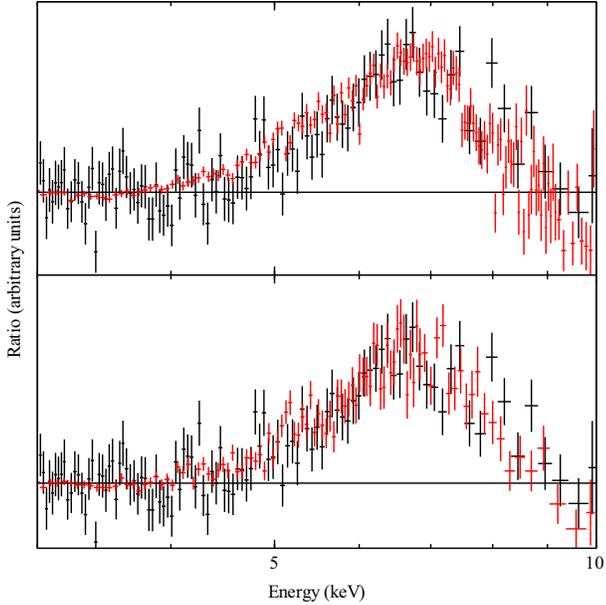}}
}
\end{center}
\caption{Comparison of the iron line profiles for the \xmm\ (black) and the latter
two \bepposax (red) observations of \xte\ (21st September and 3rd October; top and
bottom panels respectively). Again, the profiles appear to have similar breadth. A
quantitative comparison is given in Table \ref{tab_line_evol}, confirming the
similarity.}
\label{fig_1650_sax2}
\end{figure}

If the line is primarily broadened by Comptonisation in either the corona
or in a truncated disc, which moves in as the outburst progresses through the
transition from the low/hard to the high/soft state, some evolution in the 
width of the line profile is expected as gravitational effects become more
important, and the temperature of both the disc and the corona changes
(typical electron temperatures are $kT_{\rm e} \sim$ 50--100\,\kev\ in the
low/hard state and $kT_{\rm e} \sim$ 30--50\,\kev\ in the high/soft state; see
\eg \citealt{Malzac09}, \citealt{Gierlinski99}, \citealt{Sunyaev80}). In order
to determine whether any such evolution occurs we now also consider the line
profiles obtained for the latter two \bepposax observations. We model the
continuum in the same way detailed previously, with the same standard BHB
components, requiring the photon index to be in the range 1.7--3.0. In Fig.
\ref{fig_1650_sax2} we show a comparison of the \xmm\ line profile to the
profiles obtained in the second and third \bepposax observations (top and
bottom panels respectively), similar to Fig. \ref{fig_1650_sax}, although
here the strength of the iron line relative to the continuum has changed,
so the ratios have been scaled to match the continuum and the peak of the
emission. Again, the profiles appear to have a similar breadth and shape in
both cases.

\begin{figure*}
\begin{center}
\rotatebox{0}{
{\includegraphics[width=470pt]{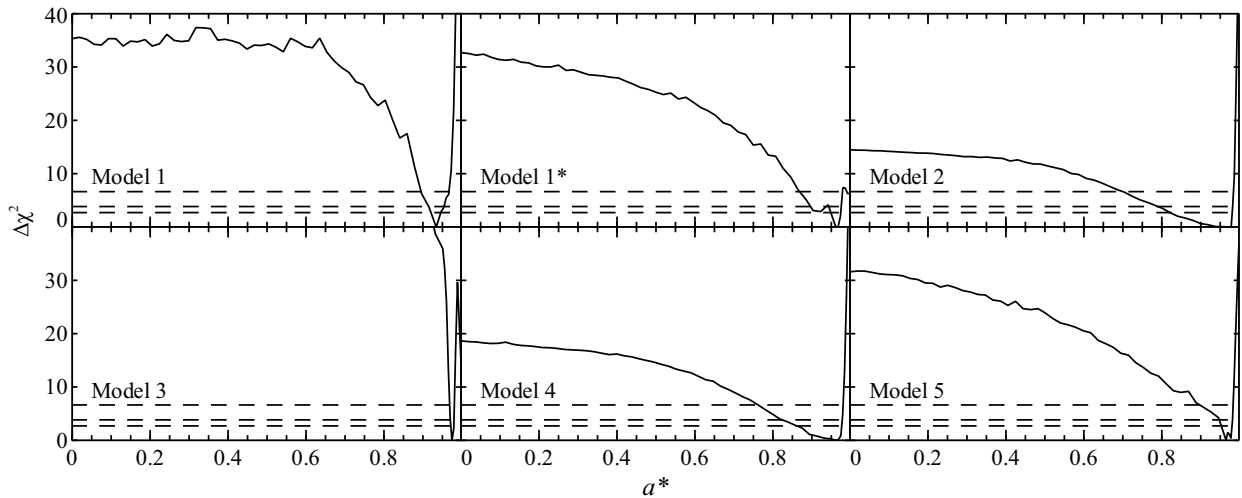}}
}
\end{center}
\caption{\chisq\ confidence contours for the spin parameter measurements presented in
Table \ref{tab_spin}, demonstrating the spin values presented in Table \ref{tab_spin}
are the global minima across the allowed range of values for $a^*$. Horizontal dashed
lines represent $\Delta$\chisq\ equivalent to the 90, 95 and 99 per cent confidence
intervals.}
\label{fig_spin}
\end{figure*}

Since in these cases the shape of the continuum is significantly different
from observation to observation, we also quantify the comparison by modelling
the iron lines from all four observations with a \laor\ profile (given that
the additional parameters included in the \laortwo\ model could not
previously be constrained we have adopted the simpler version, which assumes
a single powerlaw emissivity profile). In this case, we model all four
observations simultaneously, requiring only that the inclination does not vary
between them (although the line energies are again constrained to be between
6.4 and 6.97\,\kev). The evolution of the key parameter in determining the width of
the line, \rin, is shown in Table \ref{tab_line_evol}. Although there is some
slight variation between observations, the inner radii obtained are all very
similar, confirming the visual suggestion that there is minimal evolution of
the line profile during the progression of the outburst. This lack of evolution
strongly favours the interpretation in which the iron line arises through
reflection from the inner regions of a disc with a constant and stable inner
radius. We also note again that a similar lack of evolution is displayed by
XTE\,J1752-223 (\citealt{Reis1752}), so such behaviour is not restricted to
this individual source.

\begin{table}
  \caption{Evolution of the inner radius for the \bepposax and \xmm\ observations
of \xte\ during its 2001 outburst, obtained by modelling the iron K line line with
a \laor\ emission line profile.}
\begin{center}
\begin{tabular}{c c c c c c c}
\hline
\hline
\\[-0.25cm]
& Observation Date & & Instrument & & \rin & \\
\\[-0.2cm]
& & & & & \rg\ & \\
\\[-0.25cm]
\hline
\\[-0.25cm]
& 11th September & & \bepposax & & $1.52^{+0.02}_{-0.03}$ & \\
\\[-0.2cm]
& 13th September & & \xmm\ & & $1.63 \pm 0.04$ & \\
\\[-0.2cm]
& 21st September & & \bepposax & & $1.31 \pm 0.01$ & \\
\\[-0.2cm]
& 3rd October & & \bepposax & & $1.31^{+0.05}_{-0.04}$ & \\
\\[-0.25cm]
\hline
\hline
\end{tabular}
\label{tab_line_evol}
\end{center}
\end{table}

\subsubsection{Black Hole Spin}

As we have demonstrated that the line profile is not modified by instrumental effects, and
that relativistic disc reflection is strongly favoured over scenarios dominated by Compton
broadening, we now present a quantitative measurement of the spin of \xte, replacing \kdblur\
with the \kerrconv\ convolution model (\citealt{kerrconv}). \kerrconv\ applies the same
relativistic effects as \kdblur, assuming a broken powerlaw emissivity profile (as in equation
\ref{eqn_bpl}), but directly includes the dimensionless black hole spin as a parameter,
defined as $a^*$ = \spin, where $J$ is the angular momentum and $M$ the mass of the black
hole, so that it must be in the range [-1,1] where $a^*$ = 1 is maximal prograde rotation.
Its other parameters are the indices and the break radius of the emissivity profile, the
inclination of the disc and its inner and outer radii. In this model the inner radius is
given in terms of the inner-most stable circular orbit, and we obtain the spin of the black
hole by assuming the disc extends all the way down to this radius. We again make the
simplification of a single powerlaw emissivity profile by requiring the two indices to be
identical and assigning the break radius some arbitrary value, and the outer radius of the
disc is again set as the maximum value accepted by the model. The black hole spin, the index
of the powerlaw emissivity profile and the disc inclination were allowed to vary.

\begin{table}
  \caption{Black hole spin parameters obtained by replacing \kdblur\ with \kerrconv\ for
the five models given in Table \ref{tab_refl}. The spin measurement for the basic reflection
model is calculated both including (Model 1) and excluding (Model 1*) the 0.7--1.0\,\kev\
data, to assess the effect of the residuals on the spin measurement.}
\begin{center}
\begin{tabular}{p{0.1cm} c p{1cm} c p{0.1cm}}
\hline
\hline
\\[-0.25cm]
& & & $a^*$ & \\
\\[-0.25cm]
\hline
\\[-0.25cm]
& Model 1 & & $0.93^{+0.02}_{-0.01}$ & \\
\\[-0.2cm]
& Model 1* & & $0.96^{+0.01}_{-0.04}$ & \\
\\[-0.2cm]
& Model 2 & & $0.96^{+0.01}_{-0.12}$ & \\
\\[-0.2cm]
& Model 3 & & $0.978^{+0.002}_{-0.007}$ & \\
\\[-0.2cm]
& Model 4 & & $0.95^{+0.03}_{-0.08}$ & \\
\\[-0.2cm]
& Model 5 & & $0.97^{+0.01}_{-0.02}$ & \\
\\[-0.25cm]
\hline
\hline
\end{tabular}
\label{tab_spin}
\end{center}
\end{table}

In each case, the models including \kerrconv\ provided statistically equivalent representations
of the data as those using \kdblur, without significant differences in any of the common
parameters. The spin measurements obtained are given in Table \ref{tab_spin}, and the \chisq\
confidence contours are shown for each model in Fig. \ref{fig_spin}. We again show the spin
measurement and contour for the basic reflection model with the 0.7--1.0\,\kev\ data excluded
(Model 1*) to investigate the effect of the soft residuals on the spin value obtained with this
model. Clearly the presence of these features does not significantly modify the spin obtained,
as the values obtained including and excluding this data are consistent within their statistical
uncertainties. In all cases the black hole is rapidly rotating, but maximal spin is ruled out
at the 90 per cent level. To summarize, the overall constraint placed using \refbhb\ is $0.84
\leq a^* \leq 0.98$, although we note that if, as expected, the soft residuals are not
astrophysical the lower limit on the spin increases as only Model 1 and Model 1* then provide
valid estimates.

The spin estimates presented here are slightly higher than the value of $a^*$ = $0.79 \pm 0.01$
presented for the same dataset in \cite{Miller09}. On investigation, we find that this is
primarily due to the different reflection codes used in these works; \cite{Miller09} make use of
the \cdid\ reflection code (\citealt{cdid}), while here we use \refbhb, an adaptation of the
more recent \reflionx\ code which has specifically been calculated for use with BHBs. \reflionx\
(and hence \refbhb) includes a far more comprehensive range of ionised species for the elements
with high cosmic abundance. These two codes use different energy ranges for the incident continuum,
which may have an effect on the reflection spectrum produced. Upon replacing \refbhb\ with a
combination of \cdid\ and various multicolour disc models we also obtain lower values for the spin
of \xte, consistent with those presented in \cite{Miller09}. More recently, the reflection code
\xillver\ (\citealt{xillver}) has further extended the range of ionised species and atomic
transitions considered, but results obtained with the use of that reflection code are typically in
good agreement with those obtained with \reflionx\ when considering relativistically broadened
scenarios. Since there is no version of \xillver\ calculated specifically for use with BHBs
currently available, we do not make use of that code here.

\subsection{MCG--6-30-15}

Having demonstrated that the line profile observed in \xte\ has an astrophysical
origin, and that the data significantly prefer the interpretation in which relativistic
effects dominate, we now consider the case of the active galaxy \mcg.

\subsubsection{Continuum Modelling}

We began by modelling the spectrum of \mcg\ with the standard AGN powerlaw-like
Comptonised continuum. However, the 0.5--10.0\,\kev\ X-ray spectra of active galaxies
are rarely featureless. Type 1 AGN commonly display additional emission at soft energies
(see \eg \citealt{Gierlin04}, \citealt{Crummy06}, \citealt{Miniutti09}). In order to
phenomenologically account for this soft excess, we also included a black body emission
component. Both of these were modified by neutral absorption. As \mcg\ is an
extragalactic source, we initially allowed for both Galactic and intrinsic absorption
with two \tbabs\ components, one local component and the other at the redshift of \mcg\
($z$ = 0.007749; \citealt{Fisher95}). The Galactic column used was $3.92 \times 10^{20}$
\atpcm, based on recent 21 cm measurements (\citealt{NH}), while the intrinsic column was
free to vary. However, in the course of the modelling we found that the intrinsic neutral
component was not required, and so was removed.

As has been widely reported in previous work (see \eg \citealt{Ballantyne03b, Young05,
Miller08}; \citealt{Chiang11}), the spectrum of \mcg\ also displays clear signatures of
ionised absorption. To account for these features, we include a model for the ionised
absorber based on that of \cite{Chiang11}. In brief, this is a three-zone absorber
which also includes a treatment of the iron-L dust absorption (\citealt{Lee01}). The
absorber consists of a highly ionised, rapidly outflowing absorption zone ($\log \xi
\sim 3.82$, $v_{\rm out} \sim 1900$\,\kmps), and two moderately ionised, mildly
outflowing zones ($\log \xi \sim 2.3$ and 1.3, $v_{out} \sim 150$\,\kmps). These are
individually included using the \xstar photoionisation code, while the iron-L dust
absorption is accounted for with the inclusion of an additional \tbabs\ component. As
with \xte, the 4.0--8.0\,\kev\ data were not considered so that the continuum was
determined with minimal contribution from the iron line region. This model provides
a good representation of the remaining data, with \rchi = 1101/934.

\begin{figure}
\begin{center}
\rotatebox{0}{
{\includegraphics[width=230pt]{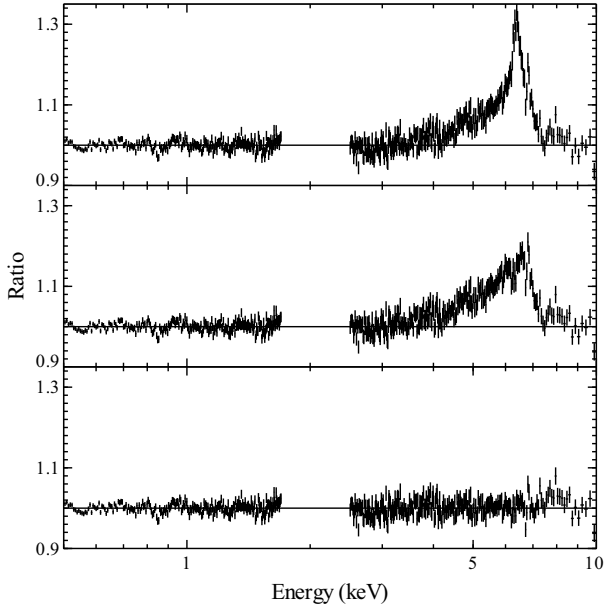}}
}
\end{center}
\caption{Data/model ratio plots for \mcg. \textit{Top panel}: ratio plot to the
continuum model fit to the 0.5--4.0 and 8.0--10.0\,\kev\ data (see text);
the combined profile of the broad and narrow iron emission components
is clearly shown. \textit{Middle panel}: ratio plot with no contribution from the
\laortwo\ line to demonstrate the profile of the broad line. \textit{Bottom panel}:
the ratio plot when both \laortwo\ and Gaussian emission lines are
included in the model; the iron emission is now clearly well modelled.
The spectra have been rebinned for clarity.}
\label{fig_mcg_line}
\end{figure}

\subsubsection{Line Profile}

As with \xte, we now include the 4.0--8.0\,\kev\ data. Fig. \ref{fig_mcg_line}
(top panel) shows the data/model ratio of the full \mcg\ 0.5--10.0\,\kev\
spectrum to the continuum model determined in the previous section. Again, it
is clear that there is a large, broad excess in the ratio spectrum in this
energy range, very similar to that seen in \xte. We again include a \laortwo\
line to model this feature as a relativistically broadened iron fluorescence
emission line from the inner regions of the accretion disc. The rest-frame
energy of the emission line was again constrained to be between 6.4 and 6.97\,\kev,
the outer radius of the disc was taken to be 400 \rg, and the rest of the line
parameters were free to vary. However, in this case, the \laortwo\ line did not
completely account for the shape of the emission feature, due to the additional
presence of a narrow core to the iron emission, which can clearly be seen in Fig.
\ref{fig_mcg_line} (top panel). This was modelled with a Gaussian emission line
at 6.4\,\kev\ (rest frame); narrow components to the iron emission in AGN can
arise due to reflection from more distant, cold material, \eg\ the dusty torus,
or due to re-emission from photoionised clouds in the broad line region. With
the inclusion of these components, the model provides a good representation of
the data, with \rchi = 1817/1725. The key parameters obtained are given in Table
\ref{tab_laor} and the data/model ratio plot is shown in Fig. \ref{fig_mcg_line}
(bottom panel). In order to demonstrate the profile of the broad iron emission,
Fig. \ref{fig_mcg_line} (middle panel) also shows the data/model ratio with no
contribution from the \laortwo\ line.

This simple analysis demonstrates the similarity of the broad excesses observed
in both the stellar mass BHB \xte\ and the AGN \mcg. In both cases these excesses
are consistent with a relativistically broadened iron emission line, originating
from the inner regions of the accretion disc around a rapidly rotating black
hole. We will now proceed to explore this similarity in more detail.

\section{Discussion and Comparison}
\label{sec_disc}

\begin{figure*}
\begin{center}
\rotatebox{0}{
{\includegraphics[width=235pt]{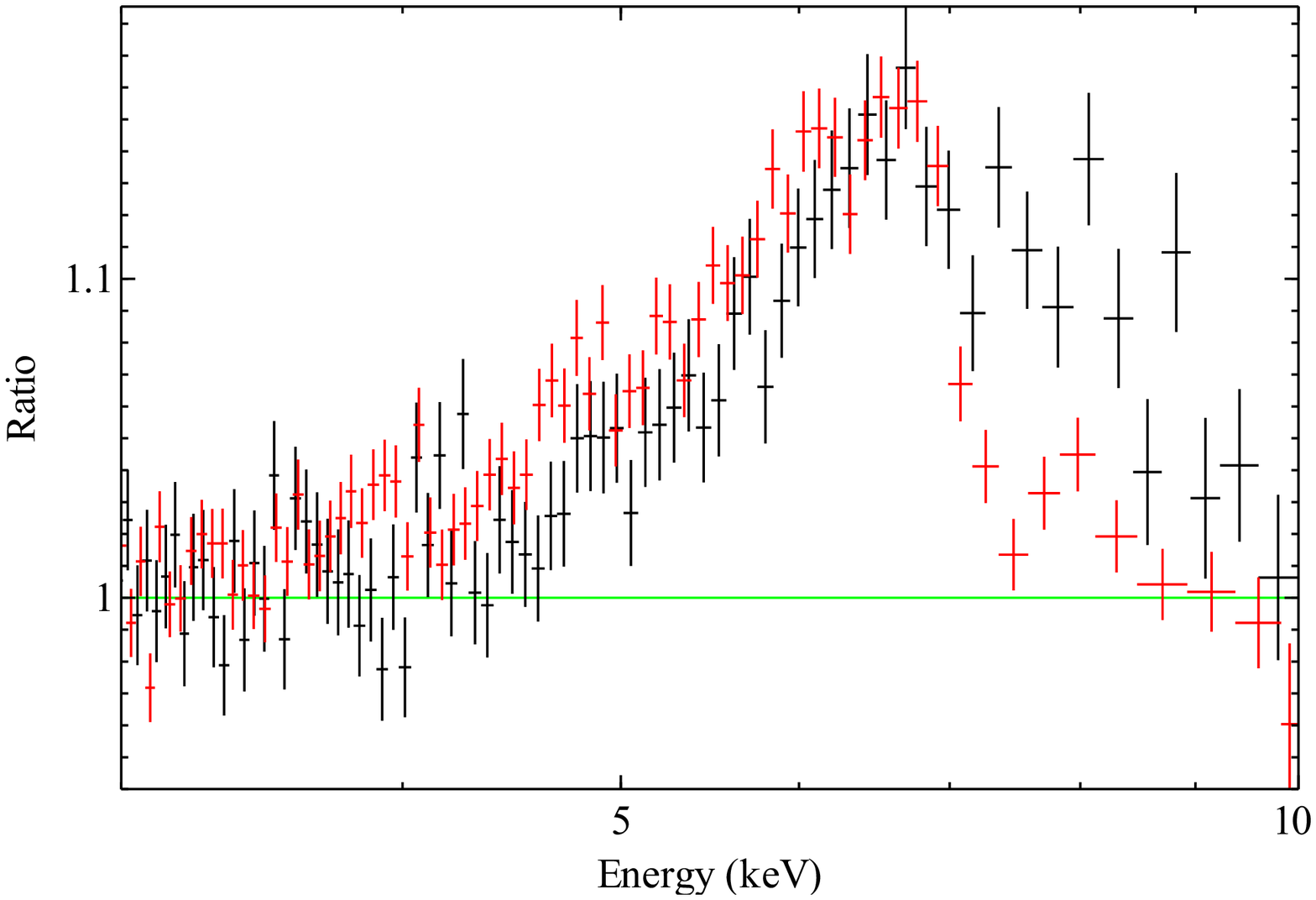}}
}
\hspace{0.5cm}
\rotatebox{0}{
{\includegraphics[width=235pt]{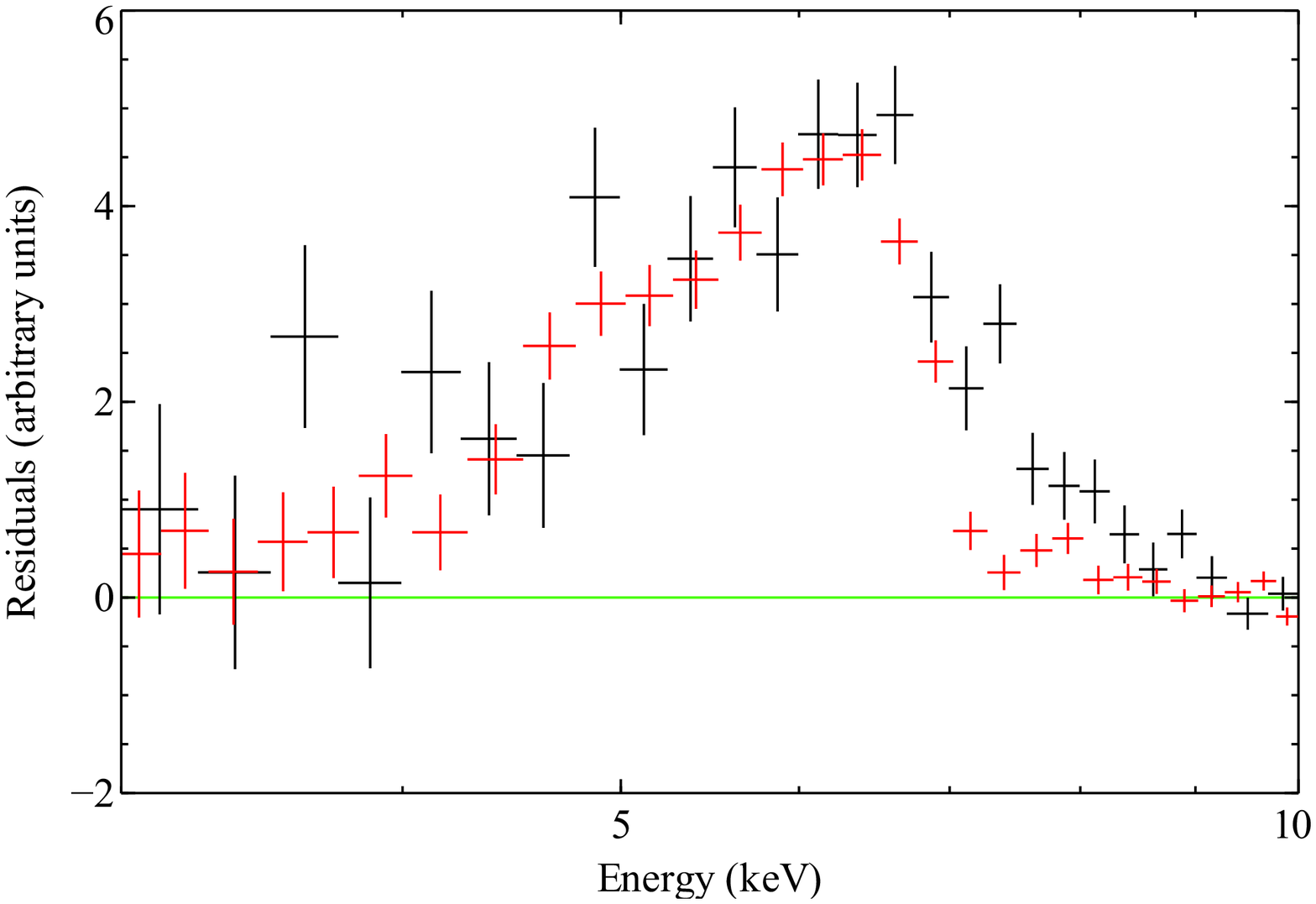}}
}
\end{center}
\caption{Data/model ratio (left panel) and Data--Model residual (right panel) plots
for the 2001 \xmm\ observations of \xte\ (black) and \mcg\ (red), highlighting the
broad iron emission lines. Residuals are in units of (normalised) cts s$^{-1}$
keV$^{-1}$; those for \mcg\ should be multiplied by 10$^{-2}$. The spectra have been
rebinned for clarity. Aside from differences in the blue wings which imply different
disc inclinations, the line profiles are quite similar.}
\label{fig_lines}
\end{figure*}

We have analysed the 2001 \xmm\ and \bepposax observations of \xte. In agreement with
previous works on these observations, we find that this source displays a broad excess
around $\sim$7\,\kev\ over the standard disc-plus-Comptonisation model for BHBs (see \eg
\citealt{Miller02b, Miniutti04, Miller09}). The profile of this excess is not particularly
dependent on the choice of continuum model. Such excesses are frequently observed in
other non-quiescent BHBs with high quality data, and are usually interpreted as iron
emission lines excited through illumination of the disc by the Comptonised continuum,
then broadened and skewed by the relativistic effects present close to the compact object.
We have confirmed that such disc reflection models can successfully reproduce the
observed data, both phenomenologically and through the use of the more physical
reflection model \refbhb\ (\citealt{refbhb}). Although the basic reflection model
provides a good representation of the broad-band shape of the spectrum, there appear to
be residual features at $\sim$0.8\,\kev\ in the \xmm\ data (see Fig. \ref{fig_xte_line}).
Through a brief analysis of the 2002 \xmm\ burst mode observations of the Crab nebula we
find that qualitatively similar residuals are present. We therefore expect that, at least
to some extent, these features are due to systematic uncertainties in either the \xmm\
burst mode calibration or in the ISM absorption model used, although in the interest of
completeness we also explore a number of possible physical origins (see section
\ref{sec_soft}). The exact origin of these residual features is not particularly important
to our broader discussion of the sources analysed in this work, as even if these residuals
are evidence for absorption by ionised material, this material does not have any
significant impact on the high energy spectrum around the iron K shell transitions (see
Fig. \ref{fig_abs1}).

In addition, we have also briefly analysed the long 2001 \xmm\ observation of \mcg. These
observations of \xte\ and \mcg\ were selected because they represent some of the highest
quality X-ray datasets available for XRBs and AGNs respectively, and previous work on these
sources invoking relativistic disc reflection has indicated that both are likely to be
rapidly rotating (see \citealt{Miller09}, and \citealt{kerrconv} for \xte\ and \mcg\
respectively). We have demonstrated that even after accounting for the ionised absorption
present in \mcg, it also appears to show a broad excess at $\sim$6--7\,\kev. Although we
limit ourselves to a phenomenological analysis of this source, this is sufficient for our
needs, and we again show that this excess can also be modelled as a relativistically
blurred iron emission line arising in the inner regions of the accretion disc due to
reflection of the intrinsic Comptonised continuum.

However, these simple demonstrations are not the main focus of this work; reflection
interpretations very similar to those constructed here have previously been proposed
for both of these sources, using both these observations and others, and have been
shown to be successful (\xte: \citealt{Miller02b, Miniutti04, Miller09}; \mcg:
\citealt{Fabian02MCG, Fabian03, Ballantyne03a, Vaughan04MCG, Reeves06, kerrconv,
Miniutti07}). In broad agreement with these previous works, our analysis indicates
the black holes powering both sources are indeed likely to be rapidly rotating,
although our quantitative spin measurements for \xte\ are typically higher than that
obtained by \cite{Miller09}. As previously stated, in the case of \mcg\ partially
covering absorption has also been shown to provide a good representation of the
observed energy spectrum and variability when allowed sufficient complexity
(\citealt{Miller08, LMiller09}). In this interpretation, the curvature at
$\sim$4\,\kev\ is due to the blue wing of a wide absorption trough rather than
relativistically broadened emission. \cite{Inoue03} and \cite{Miyakawa09} propose
similar interpretations, in which a relativistically broadened iron line is not
required, and the variability is dominated by changes in the absorbers.

Even with the quality of data available for \mcg\ the disc reflection/light bending
and partially covering/variable absorption interpretations remain statistically
indistinguishable. Instead the main focus of this work is to present a comparison
between the apparent profiles of the the excesses observed at $\sim$6--7\,\kev\ in
these two sources, which we argue favours the reflection interpretation for this
feature in AGN. We stress that this is not a proof, merely a series of logical
arguments and comparisons which we find compelling, and present here to the reader.
However, this is not a statement against complex absorption in AGNs. There is clear
evidence for ionised absorption in the example of \mcg\ used here. We are merely
arguing that this process is not the correct origin for the feature at
$\sim$6\,\kev\ interpreted here as a broad iron emission line.

It is clear from Table \ref{tab_laor} that the values for the key parameter in
determining the red wing of the iron line, \rin, obtained for \xte\ and \mcg\ are
similar. In Fig. \ref{fig_lines} we focus on the line profiles of the two sources.
A visual inspection shows that, aside from minor differences in the blue wings due
to the different inclinations at which we are observing the inner discs of these
sources, the two profiles are indeed very similar. Specifically, we draw the readers
attention to the similarity in the curvature at $\sim$4--5\,\kev, which is interpreted
here as the onset of the relativistically broadened iron emission line. This curvature
plays a strong role in determining the inner radius of the accretion disc, so after
a visual comparison it is not surprising that similar inner radii are obtained. Given
the similarity of the observed profiles in these two sources, it is natural to assume
that they arise from to the same physical process. Since the two sources considered
are separated by many orders of magnitude in terms of their black hole mass, this
process must be mass independent, and is hence likely to be atomic. Indeed, we have
already stressed that such excesses are commonly observed in AGN and BHBs. In addition,
fairly broad excesses are also observed in neutron star XRBs (see \eg \citealt{Cackett10},
\citealt{Reis09ns}, \citealt{D'ai10}, \citealt{Lin10}).

\subsection{Complex Absorption}

We have already stressed that both disc reflection and partially covering absorption
provide statistically acceptable representations of the available data for \mcg\
(in the absorption case the curvature at $\sim$4--5\,\kev\ is due to the blue
wing of an absorption trough rather than excess emission). Indeed, both of these
interpretations are atomic in nature, and should therefore be mass independent.
Given their relative proximity, Galactic XRBs are the ideal testing ground to determine
the origin of these excesses. They are bright enough that the absorption may be
very well constrained, hence we now consider the possible origins of, and observational
evidence for, absorption in XRBs, with particular reference to \xte.
It is well known that the ISM contains a combination of neutral and ionised material.
However, the Galactic column is not sufficiently high for the partially ionised species
to significantly alter the X-ray spectra of Galactic sources; this has been demonstrated
with detailed studies of a number of sightlines (\citealt{Juett04}; \citealt{Juett06}).
In addition, it is well accepted that this absorption should not vary significantly from
epoch to epoch. Considering the source itself, although the masses of the black hole and
its companion star have yet to be well constrained, the work of \cite{Orosz04} strongly
suggests \xte\ is a LMXB, and hence accretes via Roche Lobe overflow rather than through
stellar winds, so the X-ray spectrum will not suffer from obscuration due to such
outflows. Many of the other XRBs (neutron star or black hole) that show evidence for a
broad excess emission component at $\sim$6--7\,\kev\ are also LMXBs, and hence should
also be uncomplicated by strong stellar winds from their companions.

The only other major source of material that could significantly obscure the source are
outflows from the disc itself. Radiatively driven disc winds are expected to become
increasingly important as black holes approach (and possibly exceed) the Eddington Limit
(see \eg \citealt{King03}). Hydrodynamic simulations of such winds (on all scales, from
AGN accretion discs to early proto-planetary discs) suggest these outflows should most
severely modify the observed X-ray spectra at high inclination angles (\ie shallow angles
with respect to the surface of the disc), see \eg \cite{Proga00, Proga04, Proga02,
Schurch09, JOwen10, Sim10}, \etc\ The exact angle at which these winds become important
must depend on the mass and the Eddington fraction of the black hole, which will in turn
play in important role in determining the ionisation structure and the mass outflow rate
of the winds. For a $10^8$\,\msun\ AGN at an Eddington fraction of $L_{\rm E} = 0.5$,
\cite{Schurch09} find that the wind should start to appreciably modify the observed X-ray
spectra at inclinations greater than $\sim$60\deg. We obtain a similar inclination for
\xte\ yet find no evidence for a similar level of absorption, although we note that the
Eddington fraction of \xte\ in this observation is likely to be comfortably lower than
that considered in \cite{Schurch09}: assuming \mbh $\sim 5$\,\msun\ gives $L_{\rm E} \sim 0.1$. 

Orbital parameters obtained for Galactic XRBs through optical studies of the companion stars
indicate they are observed at a large range of inclination angles (see \eg \citealt{Orosz03}).
Although we do not find any strong evidence here, ionised absorption associated with outflowing
material \textit{is} seen relatively frequently in LMXBs, with notably strong examples being
GRO\,J1655-40 (\citealt{Miller06a, DiazTrigo07}) and GRS\,1915+105 (\citealt{Kotani00, Lee02,
Ueda09, Neilsen09}). As expected, this absorption is more prominent in high inclination sources,
and/or those observed at high Eddington fractions. Even at relatively high inclinations, the
absorption usually manifests itself as weak, narrow line features best studied with grating
spectrometers, although XRBs with the highest inclinations frequently show `dips' in intensity,
during which the absorption increases and, in some cases, can severely modify the spectrum (see
\eg \citealt{Sidoli02}, \citealt{Boirin05}, \citealt{DiazTrigo06}).

The general picture is that
when present, disc winds from XRBs are more ionised (at a given inclination angle) than those in
AGN. This might not be surprising as the discs around XRBs are expected to be more ionised than
their AGN counterparts, owing to their much higher temperature. Indeed, \cite{Proga02} show that
the winds in LMXBs will be thermally driven from the disc, and should in general be optically
thin. Therefore, in the vast majority of cases, these outflows do not dominate the observed
shape of the X-ray spectra of BHBs. The fact that the continuum of \xte\ presented here is well
modelled with the standard BHB components (thermal disc at soft energies and Comptonised
emission at hard energies) leads us to conclude that the X-ray spectrum of \xte\ has \textit{not}
been significantly modified by any additional complex absorption in this observation. Furthermore,
in \S \ref{sec_abs} we have demonstrated that the \xmm\ data exclude the possibility that the iron
line profile may have been modified by a strong, relativistic outflow similar to that proposed by
\cite{Done06}. We briefly note that high mass X-ray binaries (HMXBs) also frequently display fairly
strong absorption features (\eg Cygnus X-1; see \citealt{Hanke09}, \citealt{Hanke10}), but the
absorption in these sources is expected to originate due to the strong stellar winds of their
companions from which the BH accretes.

In contrast to the evidence for ionised outflows in LMXBs, although the profile of the excess
emission evolves with inclination, its presence does not. The growing picture then is that
many observations of XRBs show evidence for broad excesses at around $\sim$6--7\,\kev, despite
these sources being observed at a large range of inclinations. The vast majority of these
sources do not display any evidence for significant modification by complex absorption above
that expected due to the ISM. These excesses cannot then be associated with the presence of
disc winds, and must be intrinsic to the source.

\subsection{Common Origin}

If these excesses are not associated with complex absorption, as we have argued is highly likely,
they must arise due to some emission process. Limiting ourselves to atomic processes, as we wish
to simultaneously explain the excesses seen in XRBs and AGN, the energies at which they are
observed strongly suggest an association with iron emission. Previous works have provided detailed
comparisons of the various processes which may broaden one or more iron emission lines into an
emission feature as broad as those observed in XRBs (\eg \citealt{Hiemstra11}), and frequently
conclude that, assuming the emission features are truly as broad as they appear, it is highly
likely they have primarily been broadened by relativistic effects, which in turn requires the
intrinsic emission to arise close to the black hole. Indeed, we have considered the main alternative
for XRBs, namely that Compton broadening dominates, in some detail for the case of \xte, and have
determined that the scenario in which relativistic broadening dominates is strongly preferred (see
sections \ref{sec_alt} and \ref{sec_line_evol}). We stress again that some contribution to the line
width from Comptonisation is unavoidable in, but this is self-consistently included in the \refbhb\
reflection model primarily used here. Furthermore, as we are searching for a common origin for the
features in XRBs and AGN, we note that historically several authors have considered the possibility
that Compton broadening may be the dominant effect in producing the broad lines observed in AGN, and
have concluded that such a scenario is not viable for these sources (see \citealt{Fabian95,
Reynolds00, Ruszkowski00}).

We must therefore consider whether the excesses may be artificially broadened in some way. It has
recently been claimed that instrumental effects, such as pile-up, may contribute to the broadening
(see \citealt{Yamada09, Done10, Ng10}). However, detailed simulations by \cite{Miller_pileup} have
shown statistically that pile-up most likely has the opposite effect, and should cause a narrowing
of the features. Furthermore, we note that broad features similar to that observed here have been
detected in XRBs and AGNs over a variety of luminosity states, and with a variety of different
detectors, including gas based spectrometers which do not suffer at all from pile-up. For a recent
review on detections of broad emission lines see \cite{Miller07rev}. In the specific case presented
here, we can be confident that the spectrum of \xte\ does not suffer from pile-up effects, as the
observation was taken with \xmm\ in burst mode, which is specifically designed to observe extremely
bright sources. This has been confirmed through comparison of the line profiles obtained in this
observation and the \bepposax observation roughly a days prior; Fig. \ref{fig_1650_sax} shows the two
are practically identical, and the \bepposax MECS detectors are gas based and thus do not suffer
from pile-up. In addition, this strongly suggests that other instrumental effects, \eg CTI, are not
significantly modifying the profile of the line, as the profile appears to be independent of detector
type. Therefore, we can conclude that the width of the line has a true astrophysical origin.

All the evidence points towards the excesses in XRBs being due to relativistically-broadened iron
emission. We stress again that, even though complex absorption may initially appear to be a plausible
explanation for these features in AGN, it is not a viable solution for XRBs. Indeed, reflection of
the Comptonised emission by the accretion disc is a natural and unavoidable consequence of the
presence of both physical components to the accretion flow, as long as some fraction of this emission
is directed towards the disc, as is expected to be the case. Provided the disc extends in to the
regions of strong gravity immediately surrounding the black hole, a widely accepted geometry for black
holes with high accretion rates, the reflected emission lines will be broadened and possibly skewed by
the relativistic effects inherent in such a region. The high cosmic abundance and fluorescent yield of
iron indicate that its primary emission features, which are known to be between approximately 6.4
and 6.97\,\kev\ (rest frame), should be the most prominent.

The presence of both an optically thick accretion disc and some Comptonising region which produces
high energy photons are observationally confirmed for both XRBs and AGNs (see \eg
\citealt{Remillard06rev}, \citealt{Done07rev}, \citealt{Elvis10sed}, \etc), therefore disc reflection
does not require any additional complex emission regions over those known to be present. In addition,
gravitational light bending is a natural and unavoidable consequence for emission originating in a
region of strong gravity in the general relativistic model, so the proposed explanation for spectral
variability and steep emissivity profiles does not require any additional physics over known
gravitational effects. Therefore, we argue that both disc reflection and light bending are present in
AGNs. For the case of \mcg, further evidence in favour of the presence of disc reflection/light
bending interpretation is presented by \cite{ATurner04}, who undertake a detailed analysis of the data
obtained with the reflection grating spectrometer aboard \xmm, and find strong evidence for a variable
continuum but very little evidence for variable absorption, as would have been expected if the spectral
variability was driven by changes in the absorbers. We also note that \cite{Rossi05} find that the
relative variations of the iron line and the continuum in \xte\ favour a lightbending interpretation.

In summary, since the presence of both accretion discs and Comptonising regions in AGNs is known,
disc reflection is indeed a logical consequence, as are relativistic effects when the disc extends
close to the black hole. This is the most self-consistent interpretation for the broad excesses
observed at $\sim$\,6\kev. Given that this process is also observed in XRBs, for which
interpretations invoking complex absorption troughs are not viable, we conclude that the
fundamental emission components observed from accreting black holes are a direct thermal component
from the accretion disc (which appears in the UV for AGN), a Comptonised component extending to
high energies, and a reflection dominated component arising due to irradiation of the disc by the
Comptonised emission. Any intervening absorption must necessarily act on all these intrinsic
emission components. Other interpretations proposed for AGN add unnecessary complexities to the
theoretical framework of accretion in strong gravity, in discord with Occam's razor.

\section{Conclusions}

With a simple spectral analysis, we have demonstrated that the high quality spectra from the 2001
\xmm\ observations of the stellar mass black hole binary \xte\ and the active galaxy \mcg\ both
exhibit similar broad excesses at $\sim$4--7\,\kev. These are well modelled with relativistically-broadened
iron emission lines. Such features are commonly observed in both low mass X-ray binaries and active
galactic nuclei. However, in the case of active galaxies there is still a debate over whether this
feature arises through fluorescent iron emission or through complex absorption processes. In this
work, we highlight the similarities of the features, herein interpreted as broad iron lines, in these
two sources. By stressing that the general accretion process is mass independent, as optically thick
accretion discs and Comptonising regions are present in both classes of black hole, we argue that the
simplest solution for these apparent excesses in AGN is the same as that for their lower mass
analogues, Galactic X-ray binaries, where we hope to have convinced the reader that the presence of
relativistic disc lines is confirmed.

\section*{ACKNOWLEDGEMENTS}

The authors would like to thank the referee for providing feedback which helped improve the depth
and quality of this work. DJW and RCR acknowledge the financial support provided by STFC, and ACF
thanks the Royal Society. The authors would also like to thank F. Haberl, M. Bautz and D. M.
Walton for enlightening discussions on the effects of CTI, and M. Guainazzi for discussion and
advice on XRL and the use of \epfast. Some of the figures included in this work have been produced
with the Veusz\footnote{http://home.gna.org/veusz/} plotting package, written by Jeremy Sanders.
This work is based on observations obtained with \xmm, an ESA mission with instruments and
contributions directly funded by ESA member states and the USA (NASA).

\bibliographystyle{mnras}

\bibliography{/home/dwalton/papers/references}

\appendix
\section{Gain Fits and Broad Lines}

\begin{figure*}
\begin{center}
\rotatebox{0}{
{\includegraphics[width=230pt]{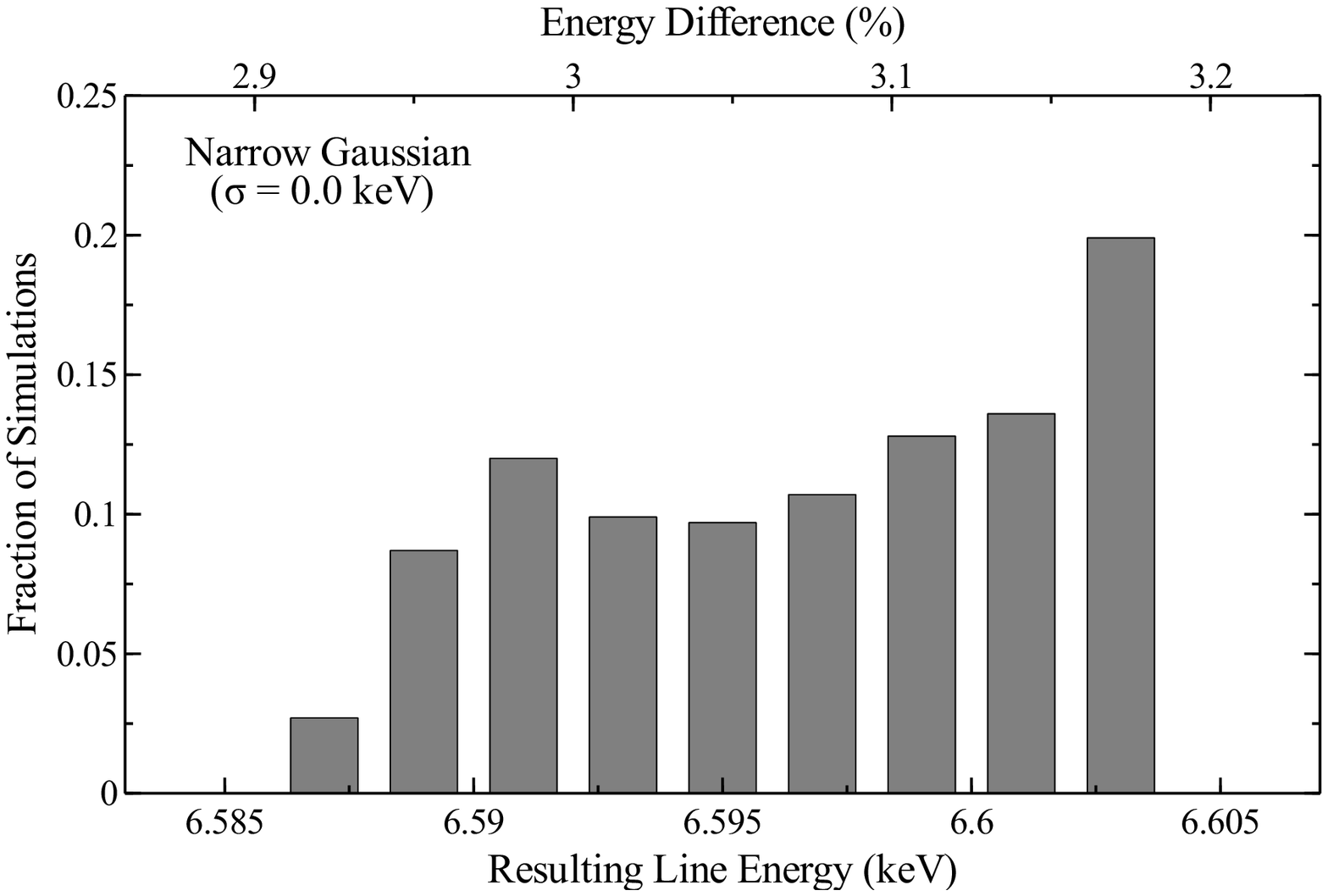}}
}
\hspace{0.75cm}
\rotatebox{0}{
{\includegraphics[width=230pt]{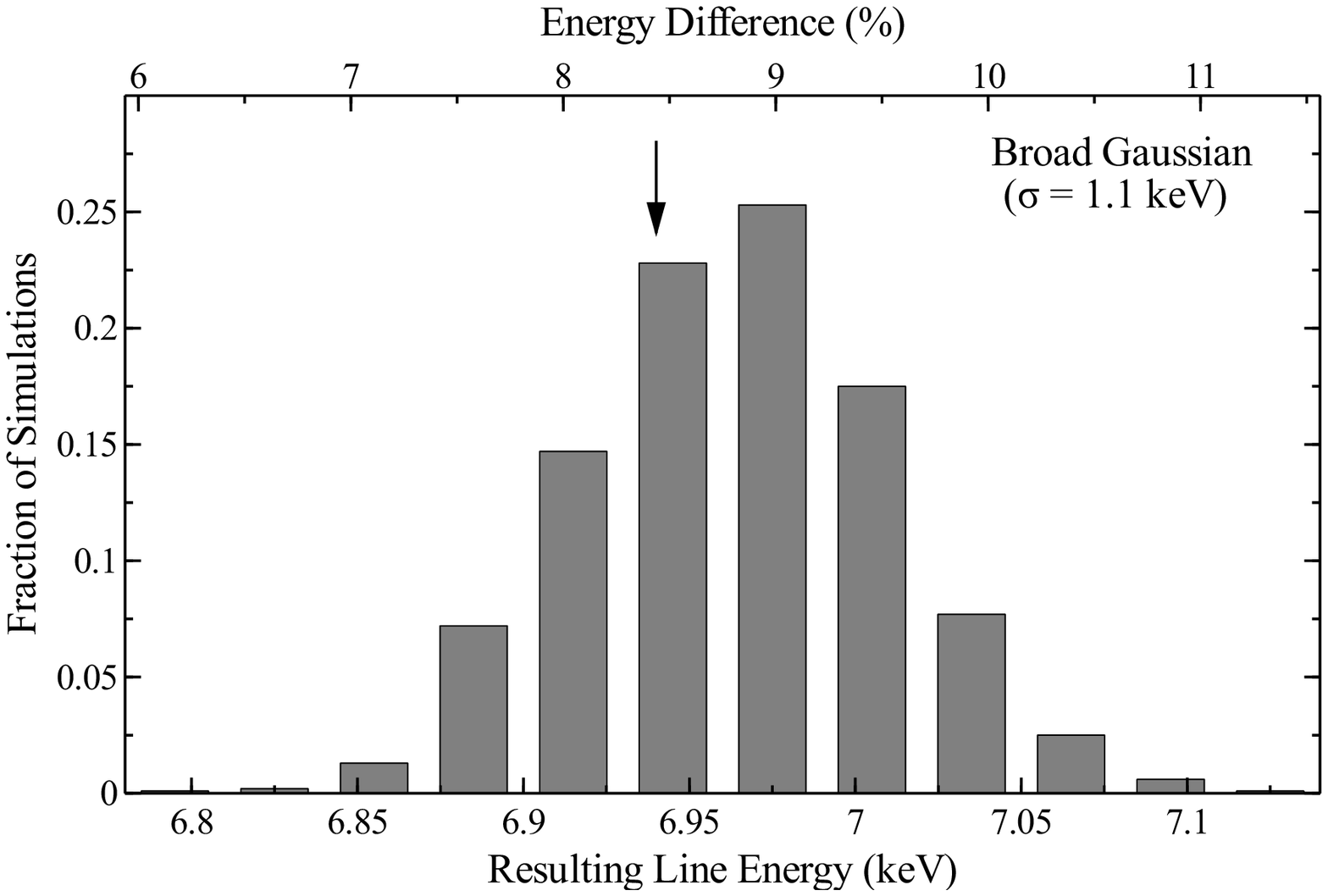}}
}
\end{center}
\vspace{-0.2cm}
\caption{Distributions of the energy centroids obtained for both narrow ($\sigma$ = 0.0\,\kev;
\textit{left panel}) and broad ($\sigma$ = 1.1\,\kev; \textit{right panel}) Gaussian line
profiles from the gain fit simulations described in the text. The lines were initially simulated
at 6.4\,\kev, and the energy centroids were measured again after the application of a 3 per cent
constant multiplicative gain shift, as applied by \epfast\ for the case of \xte. The line
centroids for the broad profile are systematically shifted by much more than the gain shift
applied. The energy shift obtained for the broad line with the \epfast-modified data in this
work is indicated with an arrow, clearly consistent with the simulated distribution.}
\label{fig_sims}
\end{figure*}

\begin{figure*}
\begin{center}
\rotatebox{0}{
{\includegraphics[width=230pt]{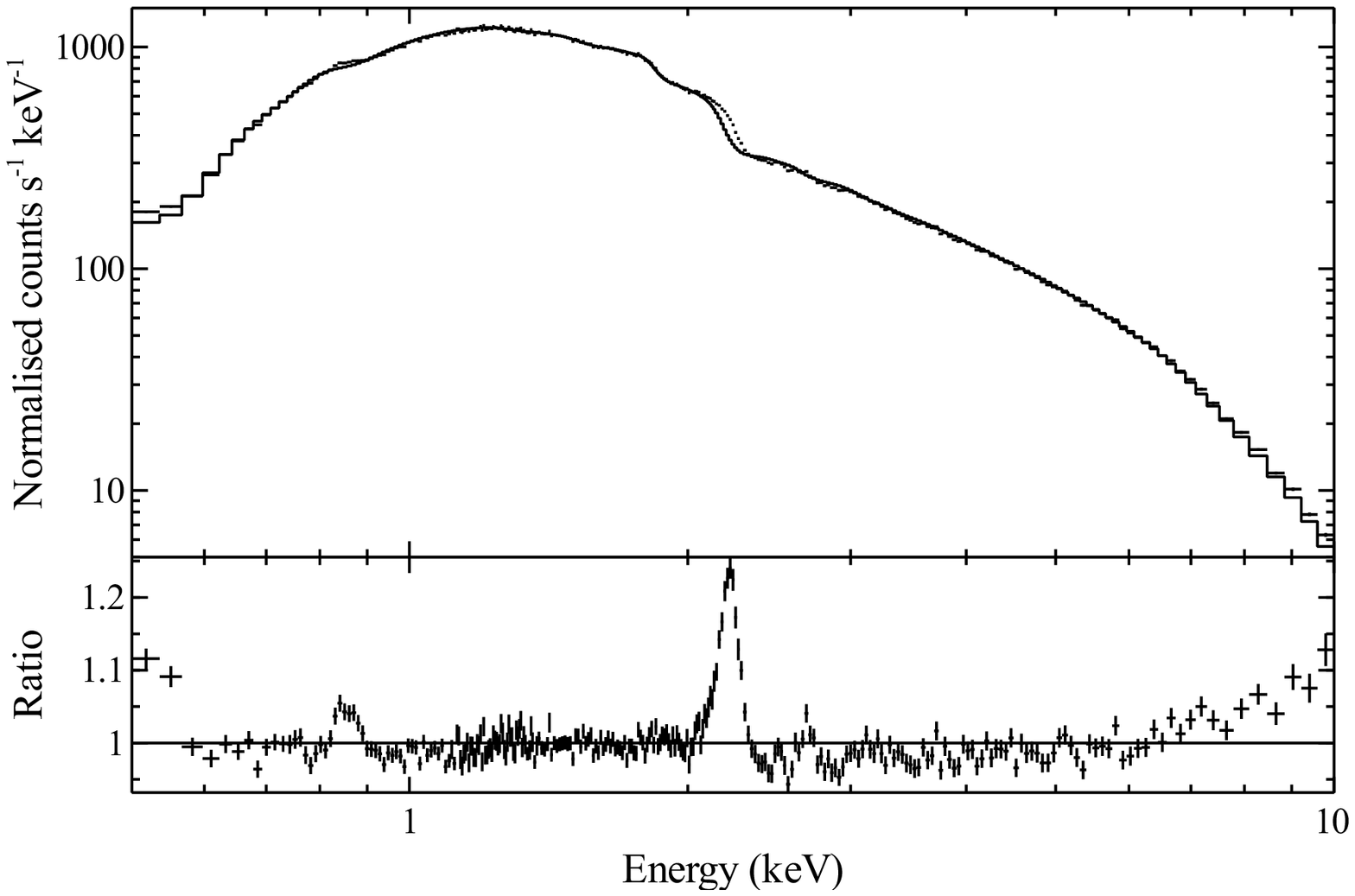}}
}
\hspace{0.75cm}
\rotatebox{0}{
{\includegraphics[width=230pt]{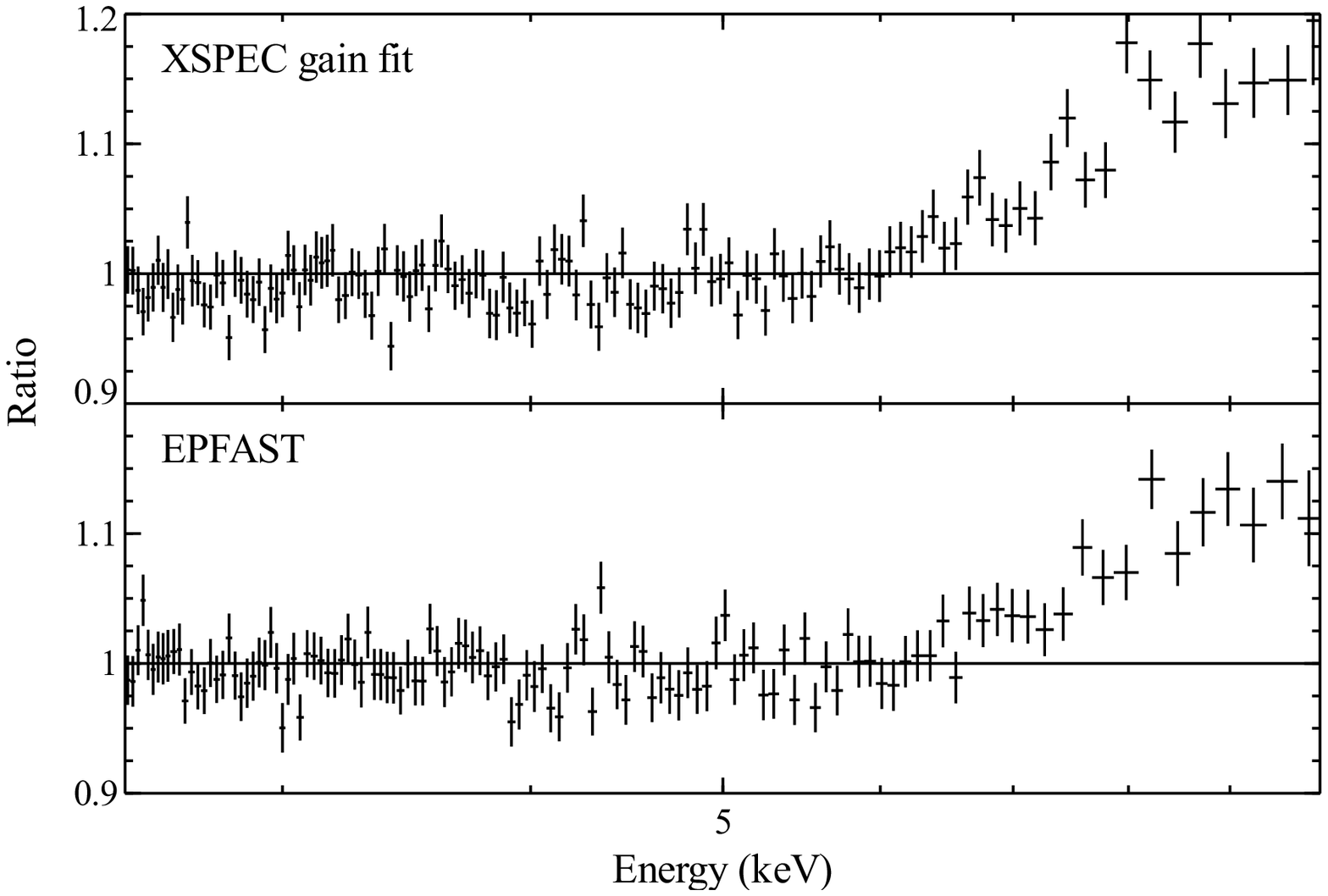}}
}
\end{center}
\vspace{-0.2cm}
\caption{\textit{Left panel}: Data simulated with an absorbed powerlaw model using the \xmm\ responses.
The same model has been fit to the data, but with a 3 per cent multiplicative gain fit applied. In
addition to the expected differences at the major absorption and instrumental edges, the gain fit
clearly results in additional, artificial curvature in the continuum at high energies. \textit{Right
panel}: A Comparison of the effect of applying EPFAST and the XSPEC `gain fit' command (see text).
The two are found to introduce identical curvature at high energies.}
\label{fig_spec_sims}
\end{figure*}

Here we present simulations demonstrating the effect of applying a constant multiplicative gain
shift on the results obtained modelling features in the iron K energy region. We simulate data
using the \epicpn\ response files from a simple model consisting of an absorbed powerlaw (\nh\
= $7\times10^{21}$\atpcm, $\Gamma = 2.15$) and a Gaussian emission line with an equivalent width
of 350\,eV at 6.4\,\kev. The exposure time for each simulated dataset was 1000\,s. We then
attempt to model this data with the same components, but with a multiplicative 3 per cent gain
shift applied to shift the response to lower energies (equivalent to shifting the data to higher
energies by the same amount, the approach taken by \epfast). These values were chosen to best
represent the case presented in the main body of the paper. This process is repeated 1000 times,
for both narrow ($\sigma$ = 0.0\,\kev) and broad ($\sigma$ = 1.1\,\kev) line profiles (line
widths were fixed throughout the modelling process, but continuum parameters could vary). In
each case, the obtained line energy was obtained, and compared with 6.4\,\kev. The results are
displayed in Fig. \ref{fig_sims}. For narrow line profiles (\textit{left panel}), the line
energies are tightly clustered around $\sim$6.6\,\kev\ ($\sim$3 per cent difference), as expected
given the applied gain shift. This is because the feature is strong and narrow, and hence changes
in the underlying continuum do not significantly effect the energy obtained.

The same is not true when the broad line profile is considered (Fig. \ref{fig_sims}, \textit{right
panel}). Even though a gain shift of 3 per cent is applied, we see that the energy obtained is
systematically shifted by a much larger amount, and the line centroid obtained is typically
$\sim$7\,\kev\ ($\sim$9 per cent shift). This is because, unlike for a narrow line, the energy
obtained for a broad line does depend somewhat on the underlying continuum model. The application
of a constant multiplicative gain shift falsely increases the high energy curvature present in
the data relative to the original continuum model; data simulated with a simple powerlaw model
display visible, concave curvature (turning up at high energies) when the response is shifted to
lower energies, see Fig. \ref{fig_spec_sims} (left panel). This additional curvature conspires to
further increase the energy centroid obtained for a broad line profile. 

\begin{figure}
\begin{center}
\rotatebox{0}{
{\includegraphics[width=230pt]{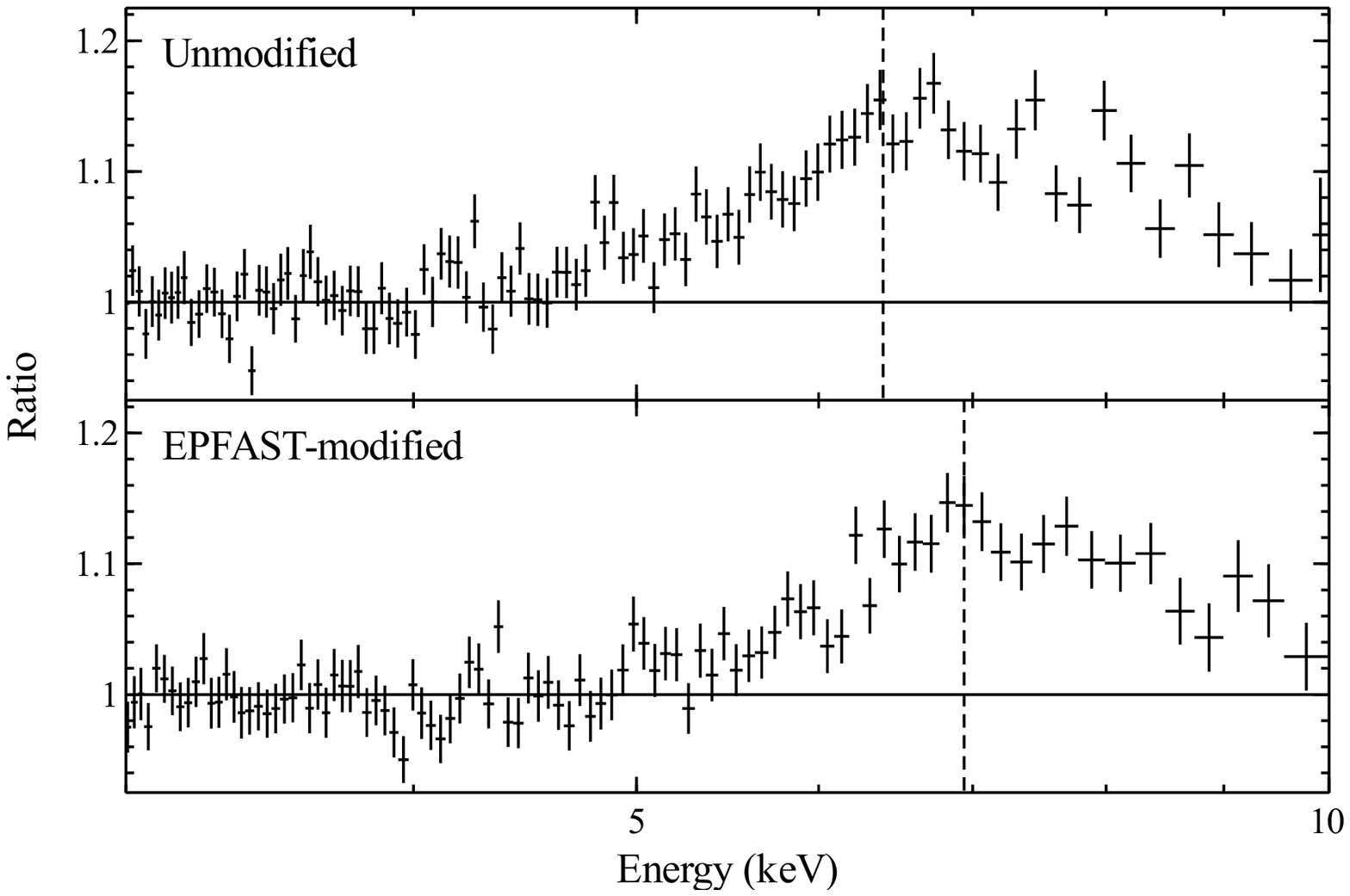}}
}
\end{center}
\vspace{-0.2cm}
\caption{The unmodified (\textit{top}) and \epfast-modified (\textit{bottom}) \xmm\ line profiles;
line energies obtained when using a Gaussian profile are shown by the dashed lines. The additional
curvature at high energies introduced by \epfast\ causes the centroid of the broad line profile to
shift by a larger amount than the tool formally applies.}
\label{fig_profiles}
\end{figure}

Finally, before we can relate this back to the shift in the line centroid energy observed when
applying \epfast, we must first confirm that \epfast\ and the `gain fit' command in \xspec\ used
in the above simulations do indeed have the same effect at high energies. To do this, we apply
the basic reflection interpretation (model 1) to the unmodified \xmm\ data with the 3 per cent
gain shift used in the above simulations. Only the overall normalisation of the model is free to
vary, all other parameters are fixed at the values in Table \ref{tab_refl}. Examining the
data/model ratio (Fig. \ref{fig_spec_sims}; right panel, top), the same high energy curvature as
highlighted previously can be seen. Next we apply this model (with no gain shift) to the
\epfast-modified data, again only allowing the overall normalisation to vary. Examining the
data/model ratio here (Fig. \ref{fig_spec_sims}; right panel, bottom), we see identical curvature.
It is clear from this comparison that the use of \epfast\ has the same effect at high energies
as the \xspec\ gain fit command, also introducing additional curvature into the high energy
continuum. We therefore conclude that the shift of $\sim$0.55\,\kev\ in the line energy obtained
with the \epfast-modified data in the main body of this paper, highlighted again in Fig.
\ref{fig_profiles}, has been strongly amplified by the incorrect influence of \epfast\ on the
continuum, and is a natural consequence of its application in this instance.

\label{lastpage}

\end{document}